\documentclass[twocolumn]{aastex701}

\newcommand{\Msun}{M$_{\odot}$}
\newcommand{\Mbh}{$M_{\rm BH}$}

\newcommand{\Mstar}{$M_{\star}$}

\newcommand{\Lsun}{L$_\odot$}

\newcommand{\ml}{\emph{M/L}}

\newcommand{\hst}{\emph{HST}}
\newcommand{\kms}{km~s$^{-1}$}

\newcommand{\go}{G$_{\rm outer}$}

\newcommand{\cotwo}{$^{12} $CO(2$-$1)}

\newcommand{\siggas}{$\sigma_{\rm gas}$}
\newcommand{\coone}{$^{12} $CO(1$-$0)}
\newcommand{\kinms}{\textsc{KinMS}}
\newcommand{\skysampler}{\textsc{SkySampler}}

\hypersetup{colorlinks, linkcolor=blue, citecolor=blue, urlcolor=blue}

\usepackage{multirow}
\usepackage{footnote}
\usepackage{newtxtext,newtxmath}
\usepackage{graphicx}	
\usepackage{amsmath}	
\usepackage{caption}
\usepackage{xcolor}

\graphicspath{{./}{figures/}}

\received{June 9, 2026}
\revised{October 6, 2026}
\submitjournal{ApJ}

\shorttitle{Molecular-based SMBH measurement of NGC 3862}
\shortauthors{T.\ N.\ Le et al.}
\begin{document}
\title{\large{\bf Unveiling the Central Black Hole of NGC 3862: A Direct Mass Determination Using ALMA’s High-Resolution Molecular Gas Kinematics}}

\correspondingauthor{Dieu D. Nguyen}\email{dieun@umich.edu}

\author[0009-0009-0015-1208]{Tuan N. Le}
\email{tuan.le.nutshell@gmail.com}
\affiliation{Faculty of Physics—Engineering Physics, University of Science, VNU-HCM, Ho Chi Minh City, Vietnam}
\affiliation{Viet Nam National University Ho Chi Minh City, Vietnam}
\author[0000-0002-5678-1008]{Dieu D.\ Nguyen}
\email{nddieuphys@gmail.com, dieun@umich.edu}
\affiliation{Department of Astronomy, University of Michigan, 1085 South University Avenue, Ann Arbor, MI 48109, USA}
\author[0009-0006-5852-4538]{Hai N.\ Ngo}
\affiliation{Department of Astronomy, University of Michigan, 1085 South University Avenue, Ann Arbor, MI 48109, USA}
\email{hai10hoalk@gmail.com}
\author[0000-0001-5802-6041]{Elena Gallo}
\email{egallo@umich.edu}
\affiliation{Department of Astronomy, University of Michigan, 1085 South University Avenue, Ann Arbor, MI 48109, USA}
\author[0009-0004-3689-8577]{Tinh Q.\  T.\ Le}
\email{lethongquoctinh01@gmail.com}
\affiliation{Department of Physics, International University, Vietnam National University in Ho Chi Minh City, Vietnam}
\affiliation{Viet Nam National University Ho Chi Minh City, Vietnam}
\author[0009-0005-8845-9725]{Tien  H.\  T.\ Ho}
\email{htien2808@gmail.com}
\email{tonggiahuy191203@gmail.com}
\affiliation{Faculty of Physics—Engineering Physics, University of Science, VNU-HCM, Ho Chi Minh City, Vietnam}
\affiliation{Viet Nam National University Ho Chi Minh City, Vietnam}
\author[0000-0002-4436-6923]{Fan Zou}
\email{fanzou@umich.edu}
\affiliation{Department of Astronomy, University of Michigan, 1085 South University Avenue, Ann Arbor, MI 48109, USA}
\author[0000-0001-9649-2449]{Que T.\ Le}
\email{ltque@hcmiu.edu.vn}
\affiliation{Department of Physics, International University, Vietnam National University in Ho Chi Minh City, Vietnam}
\affiliation{Viet Nam National University Ho Chi Minh City, Vietnam}
\author[0009-0008-6050-5736]{Huy G.\ Tong}
\email{tonggiahuy191203@gmail.com}
\affiliation{Institute of Astronomy, National Tsing Hua University, Kuang Fu Rd. Sec. 2, Hsinchu, 30013, Taiwan}
\author[0009-0005-1314-9992]{Huy M.\ B.\ Tran}
\email{tran.ba@ufl.edu}
\affiliation{Department of Astronomy, University of Florida, P.O. Box 112055, Gainesville, FL32611, USA}

\begin{abstract}
We present the direct dynamical measurement of the supermassive black hole (SMBH) mass in NGC~3862 (3C~264), a massive Fanaroff--Riley type~I radio galaxy. Using ALMA \cotwo\ observations, we map the cold molecular gas kinematics in the circumnuclear disk and model them with the \kinms\ forward-modeling code. The \cotwo\ disk exhibits a nearly face-on, ring-like morphology with regular disk rotation and a steep central velocity rise, providing clean dynamical constraints on the central gravitational potential. Our best-fit model yields $M_{\rm BH} = 2.40^{+0.52}_{-0.32}[{\rm stat.}] \pm 0.91 [{\rm sys.}] \times 10^9$~\Msun\ and a stellar mass-to-light ratio of \ml$_{\rm F814W} = 4.11^{+0.99}_{-0.27}[{\rm stat.}] \pm 0.17 [{\rm sys.}]$~\Msun/\Lsun\ (both at $1\sigma$ confidence level). This dynamically derived \ml$_{\rm F814W}$ consistent with an old, metal-rich stellar population and a Salpeter IMF. The SMBH sphere of influence is spatially resolved, confirming the robustness of our measurement.  This places NGC~3862 among SMBHs with direct molecular gas dynamical constraints, contributing to the census needed to calibrate SMBH--host galaxy co-evolution models at the high-mass end of the scaling relations.
\end{abstract}



\section{Introduction}\label{sec:intro}

\begin{figure*}[!th]
    \centering
    \includegraphics[width=2\columnwidth]{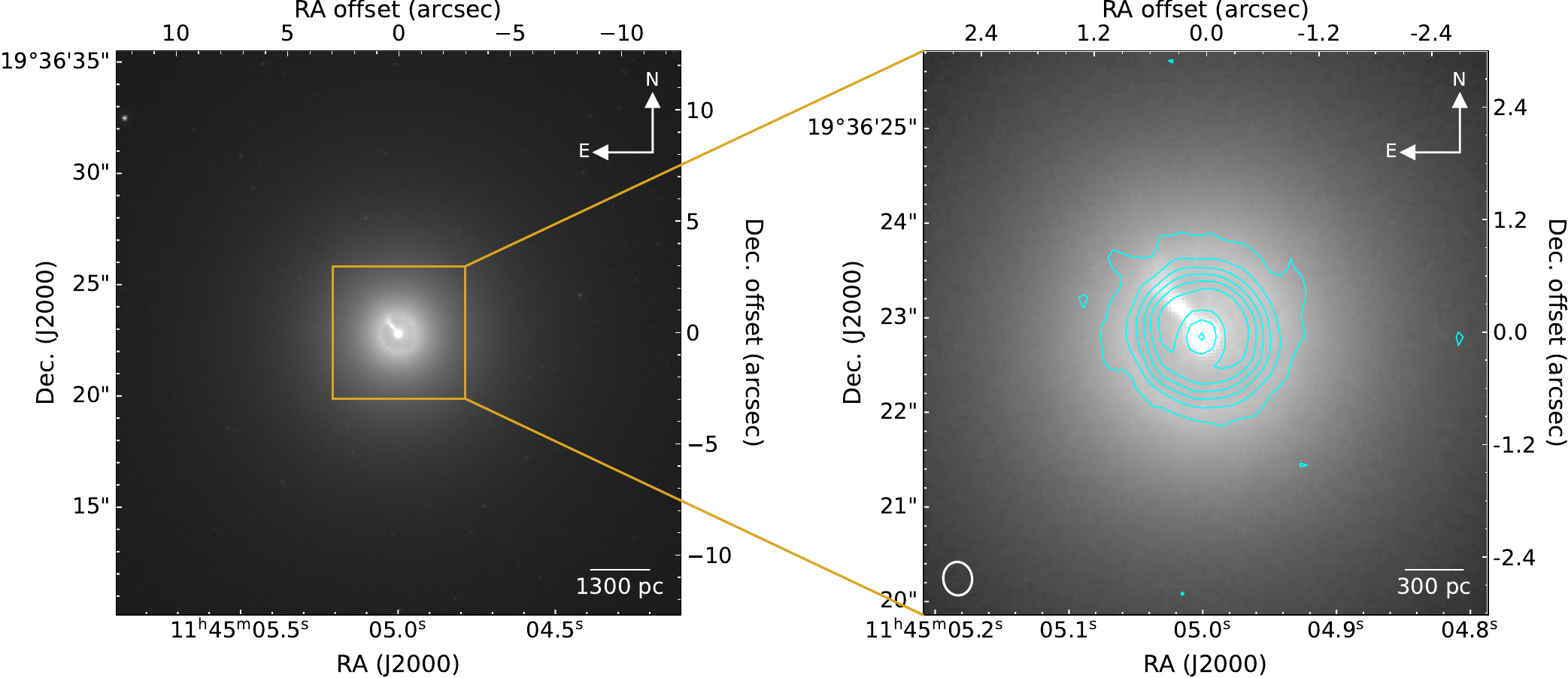}
    \caption{{\it Left:} \hst/WFC3 F814W image of NGC~3862 centered on the nucleus, displayed over a $25.6\arcsec \times 25.6\arcsec$ ($12.34~{\rm kpc} \times 12.34~{\rm kpc}$) field of view. The circumnuclear dust disk is clearly visible as a dark absorption feature against the stellar continuum. North is up, and east is to the left. {\it Right:} Zoom-in of the central $6.0\arcsec \times 6.0\arcsec$ ($2.9~{\rm kpc} \times 2.9~{\rm kpc}$) region (yellow box in the left panel), with ALMA \cotwo\ integrated intensity (moment-0) contours overlaid in cyan. Contours begin at 5$\sigma$ and increase in steps of $13\sigma$, where $1\sigma = 16~{\rm mJy~beam}^{-1}~{\rm km~s}^{-1}$. The ALMA synthesized beam ($0.348\arcsec \times 0.293\arcsec$, PA $= 6.1\arcdeg$) is shown as a white ellipse in the lower-left corner. The \cotwo-molecular gas emission is spatially coincident with the dust disk, consistent with a rotating circumnuclear disk.}
    \label{fig:HST-overlaid}
\end{figure*}

\begin{table}
\raggedright
\caption{Properties of NGC~3862.} \label{tab:properties_NGC3862}
\footnotesize
\begin{tabular}{lcr}
\hline
\hline
Parameter (Unit)                           & Value   & Ref. \\
(1)                                        & (2)     & (3)  \\
\hline
\hline
Morphology                                 & E       & \\
Nuclear activity                           & LLAGN   & \\
Right ascension (R.A.)                     & 11$^{\rm h}$45$^{\rm m}$05.01$^{\rm s}$ & \\
Declination (Decl.)                        & +19$^\circ$36$'$22.74$''$ & \\
Position angle (PA, $^\circ$)              & 148.2   & (1) \\
Ellipticity ($\epsilon$)                   & 0.06    & (1) \\
Distance ($D$, Mpc)                        & 99.4    & (2) \\
Redshift                                   & 0.0217  & (3) \\
Physical scale ($\text{pc arcsec}^{-1}$)   & 482     &     \\
Effective radius (arcsec)                  & 11.1    & (4) \\
Ellipticity ($\epsilon$)                   & 0.08    & (4) \\
$K$-band absolute mag ($M_{K}$, mag)       & $-25.5$ & (5) \\
Spin parameter $\gamma_{e}$                & $0.045$ & (5) \\
Total stellar mass (\Mstar, \Msun)         &$4.79\times10^{11}$  & (5) \\
Central velocity dispersion ($\sigma_c$, km/s) & 248 & (5) \\
Velocity dispersion ($\langle\sigma_\star\rangle_e$, km/s) & 228 & (5) \\
$\log L_{\rm peak}$ (erg s$^{-1}$)         & 43.77   & (6) \\
$\log L_{5\,{\rm GHz}}$ (erg s$^{-1}$)     & 39.98   & (7) \\
$\log L_{2-10\,{\rm keV}}$ (erg s$^{-1}$)  & 41.87   & (7) \\
$\log L_{[\rm O\,III]}$ (erg s$^{-1}$)     & 39.2    & (7) \\
Eddington-scaled luminosity $\mathcal{R}$  & $6.0 \times 10^{-5}$ & (7) \\
Jet velocity fraction                      & $\sim 0.98\,c$  & (8) \\ 
\hline
\end{tabular}
\tablecomments{(1): \cite{Ene_2018_MASSIVE_X}; (2): \cite{Ma_2014}; (3): \cite{Casadei2024}; (4): \cite{Matthew24}; (5): \cite{Veale_2017}; (6): \cite{Meyer_2011} ; (7): \cite{Torresi2018}; (8): \cite{Meyer2015} }
\end{table}

Observations over five decades have established that supermassive black holes  (SMBHs; $M_{\rm BH} \approx 10^{6}$--$10^{9}$~\Msun) are ubiquitous components at the centers of massive galaxies \citep[e.g.,][]{LyndenBell69, Kormendy13}. Despite representing only a small fraction ($\lesssim$0.1\%) of their hosts' total stellar mass \citep[e.g.,][]{Reines_2015}, SMBHs exhibit tight empirical correlations with global galaxy properties, including stellar velocity dispersion ($\sigma_\star$), bulge luminosity ($L_{\rm bulge}$), and stellar mass ($M_\star$) \citep[e.g.,][]{ferrarese2000fundamental, Saglia16, Sahu19a}. These scaling relations are widely interpreted as evidence for the co-evolution of SMBHs and their host galaxies, likely regulated by active galactic nucleus (AGN) feedback that couples black hole (BH) growth to the suppression of star formation \citep[e.g.,][]{Heckman2014_Co_evolu, McNamara12}.

However, the universality of these relations remains actively debated. The slopes of SMBH, galaxy correlations differ significantly between the low-mass \citep[e.g.,][]{Nguyen_2017, Nguyen_2018, Nguyen_2019, Greene2020} and high-mass \citep[e.g.,][]{Hlavacek-Larrondo2012, McConnell2011, McConnell2012, Walsh_2015, Walsh_2016, Walsh_2017, Thomas2016, Mehrgan2019, Nightingale2023, Melo-Carneiro2025} regimes. At the high-mass end the existence of ``ultramassive'' BHs ($M_{\rm BH} \gtrsim 10^{10}$~\Msun) and the increased scatter in scaling relations suggest distinct evolutionary pathways, such as dry mergers \citep[e.g.,][]{Boylan-Kolchin2006, Naab2006} or sustained gas accretion \citep[e.g.,][]{Silk1998, DiMatteo_2008, Fabian2012}, that are not captured by standard co-evolution models. This tension is further compounded by the recent {\it James Webb Space Telescope} \citep[{\it JWST};][]{Gardner2006} discovery of overmassive BHs in high-redshift ($z > 4$) galaxies \citep[e.g.,][]{Pacucci_2024, Greene_2024}, which implies that deviations from the local scaling relations may already be established at early cosmic times. Direct, accurate dynamical \Mbh\ measurements spanning a wide range of galaxy types and mass regimes are therefore essential for anchoring these relations, testing co-evolution scenarios, and identifying the physical processes governing galaxy--SMBH co-assembly \citep[e.g.,][]{Graham2013, Cappellari_2016, Krajnovic_2018, de_Nicola_2025}.

Direct \Mbh\ determinations require resolving the gravitational sphere of influence \citep[SOI;][]{Merritt2004}\footnote{The region surrounding the BH within which its gravitational potential dominates the stellar dynamics, defined by $r_{\rm SOI} = GM_{\rm BH}/\sigma_\star^2$.}. Water megamaser emission remains the ``gold standard'' for geometric precision \citep{Miyoshi1995}, but is restricted to rare, suitably oriented disk systems \citep{Lo_2005, Kuo_2011}. For the broader galaxy population, this condition can be met through dynamical modeling of stellar \citep[e.g.,][]{Nguyen2014, Ahn_2018, Thater_2022, Thater_2023, Nguyen_2025_NSC, Nguyen_2025_4736, Voggel_2018} and gaseous \citep[e.g.,][]{Onishi_2015, Boizelle_2021} kinematics, which have now yielded a substantial census of direct \Mbh\ measurements across a wide range of galaxy types \citep{McConnell_2013, Kormendy13}. Nevertheless, expanding this census in the high-mass regime remains essential for calibrating BH--host scaling relations \citep[e.g.,][]{Nguyen_2023, Nguyen_2026_highz}.

Although stellar dynamical modeling has long served as the primary workhorse for direct \Mbh\ measurements across a wide range of galaxy types \citep{Bosch_2016}, it is subject to several systematic limitations: sensitivity to dust obscuration \citep{Thater2017, Thater2019}, contamination from AGN continuum emission \citep{Nguyen_2026_M81}, and the mass--anisotropy degeneracy inherent in orbit-based modeling \citep{Krajnovic_2018, Nguyen_2026_4258}. Ionized-gas kinematics offer an observationally efficient alternative, but are critically susceptible to non-circular motions, which can substantially bias or even invalidate the inferred \Mbh\ \citep{Sarzi_2001, Neumayer_2007}. These limitations motivate the use of dynamically cold molecular gas as a more robust kinematic tracer.

In recent years, cold molecular gas dynamics has emerged as a powerful complement to stellar dynamics for direct \Mbh\ measurements \citep[e.g.,][]{Ruffa_2019, Kabasares_2022, Cohn_2023_PGC_1117, Kabasares2024_NGC4786_NGC5193, cohn_2024, Dominiak_2024_4751, Nguyen_2026_4061, Zhang_Hengyue_2025_1574, Zhang_Hengyue_2025_383}. Cold molecular gas frequently settles into geometrically thin, circumnuclear disks (CNDs) deep within the SOI, where it follows nearly circular orbits that directly trace the central gravitational potential. This approach significantly reduces the systematic uncertainties inherent in stellar dynamical modeling, most notably those arising from the stellar mass-to-light ratio \citep[\ml;][]{Nguyen_2026_4258, Nguyen_2026_M81} and orbital velocity anisotropy \citep[e.g.,][]{Taylor_2025, Tahmasebzadeh_2025}.

In this paper, we present the first direct measurement of \Mbh\ in NGC~3862 using high-spatial-resolution \cotwo\ observations obtained with the Atacama Large Millimeter/submillimeter Array (ALMA) in Cycle~7. We model the cold molecular gas kinematics using the KINematic Molecular Simulation (\kinms\footnote{\url{https://github.com/TimothyADavis/KinMSpy}.}) package \citep{Davis_2013, Davis_2014}, which forward-models observations of arbitrary gas distributions from interferometric and integral field unit data.

NGC~3862 is a massive FR\,I radio galaxy with an early-type (elliptical) morphology, a stellar mass of $M_\star = 4.79\times10^{11}$~\Msun\ \citep{Veale_2017}, an effective radius of $R_e = 11\farcs1$ \citep{Ma_2014}, and a luminosity-weighted stellar velocity dispersion within $R_e$ of $\langle\sigma_\star\rangle_e = 228$~\kms\ \citep{Veale_2017}. It is located at $z = 0.0217$ \citep{Casadei2024}, corresponding to a distance of $D = 99.4$~Mpc \citep{Ma_2014}. The basic properties of NGC~3862 are summarized in Table~\ref{tab:properties_NGC3862}. {\it HST} imaging in the optical and NIR reveals a well-defined, nearly face-on circumnuclear dust disk (CND) with a projected radius of $R_{\rm dust} = 0.38$~kpc, semi-axes of $(b,\,a)_{\rm dust} = (0\farcs80,\,0\farcs84)$, and a position angle of $\Gamma_{\rm dust} = -9\fdg0$ \citep{Davidson_2024}, which aligns with the nuclear molecular gas disk (see Figure~\ref{fig:HST-overlaid}), providing a compelling case for high-spatial-resolution gas-dynamical modeling to deliver such a measurement.

NGC~3862 hosts the radio source 3C~264---the fourth non-blazar AGN detected at very high energies (VHE; $>$100~GeV), discovered by VERITAS in 2018 and the most distant radio galaxy detected in the VHE band to date \citep{Mukherjee2018, Archer2020_VHE}. It bears a strong resemblance to M87---also a VHE FR\,I radio galaxy---sharing a prominent one-sided jet with superluminal apparent motions of $\sim$$7c$ \citep{Meyer2015}, comparable to the $\sim$$6c$ measured in M87 \citep{Biretta_1999}. Multiwavelength observations reveal unambiguous LLAGN activity: radio core, hard X-ray, synchrotron peak, and narrow-line [O\,\textsc{iii}] luminosities, together with an Eddington-scaled accretion rate of $\mathcal{R} = 6.0\times10^{-5}$ \citep{Torresi2018, Meyer_2011}, confirming NGC~3862 as an ideal laboratory for probing the connection between \Mbh\ and jet production in the low-accretion regime.

To date, no direct dynamical \Mbh\ measurement exists for NGC~3862. The only available estimate is an upper limit of \Mbh\ $\leq 6\times10^8$ \Msun\ reported by \citet{Beifiori_2012}, derived from HST/STIS nebular emission-line widths assuming an unresolved Keplerian disc with inclination $i = 33\arcdeg$. While low-resolution CO emission has been detected in the nucleus \citep{Davidson_2024}, no spatially resolved kinematic modeling has been performed to constrain the central gravitational potential. This work presents the first such measurement.

The paper is organized as follows. Section~\ref{sec:alma_data} describes the ALMA observations, data reduction, and analysis of the \cotwo\ kinematics. Section~\ref{sec:stellar-mass} presents the stellar mass modeling of NGC~3862 based on \hst\ imaging. Section~\ref{sec:dynamical-models} details the dynamical modeling, results, and uncertainty sources. Section~\ref{sec:conclusion} summarizes our conclusions. Throughout this paper, we adopt a flat $\Lambda$CDM cosmology with $H_0 = 70$~\kms~Mpc$^{-1}$, $\Omega_{\Lambda} = 0.7$, and $\Omega_{\rm m} = 0.3$, yielding a physical scale of $1\arcsec \approx 482$~pc at the distance of NGC~3862.

\begin{table*}[!th]
	\caption{Properties of ALMA observing tracks.}
    \hspace{-12mm}
	\label{tab:Observing Tracks}
	\begin{tabular}{cccccccccc}
	\hline\hline
Project code&Obs. tracks&Obs. date&Config.&Baseline range&ToS&   MRS   &$\theta_{\rm FWHM}$&Calibration\\
            &     &         &       &              &(seconds)&(arcsec)&(arcsec)&Pipeline\\
     (1)    & (2) &    (3)  &  (4)  &      (5)     &   (6) &     (7)     & (8) &  (9)    \\ 
    \hline
    \multirow{2}{*}{2019.1.00036.S} 
    & {\tt uid\_A002\_Xea90c0\_X47c9} & 2021-03-30 & C-5 & 15.0~m--1.3~km & 3162 
    & \multirow{2}{*}{4.9} & \multirow{2}{*}{0.3} & \multirow{2}{*}{\textsc{CASA}~6.1.1.15} \\
    & {\tt uid\_A002\_Xea90c0\_X48c2} & 2021-03-30 & C-5 & 15.0~m--1.3~km & 3161 
    & & & \\
    \hline
	\end{tabular}
    \tablecomments{(2) Observational track ID. (3) Observational date (year-month-day). (4) ALMA configuration array. (5) Minimum and maximum baseline length. (6) Total on-source time (ToS). (7) Maximum recoverable scale (MRS), i.e.\ the largest angular scale that can be recovered with the given array. (8) the average synthesized beam size. (9) Calibration method.} 
\end{table*}

\begin{table*}
	\caption{Properties of spectral windows of adopted observing tracks.}
    \hspace{-12mm}
	\label{tab:Spectral windows}
	\begin{tabular}{ccccccccc}
		\hline\hline
Project code & Obs. track & SPW & Target &Freq. range&Cen. freq.&Flagged&Bandpass& Phase\\
             &       &    &(Line)&     (GHz) &  (GHz)  &    & (Quasars) & (Quasars)\\
         (1) &  (2)  & (3) &    (4)    &          (5)         &     (6)     &       (7)       &(8) & (9)\\ 
\hline                                                           
\multirow{4}{*}{2019.1.00036.S}&\multirow{4}{*}{{\tt uid\_A002\_Xea90c0\_X47c9}} &0& cont. &(222.70, 224.58) &223.6 &No &\multirow{4}{*}{J1058+0133}&\multirow{4}{*}{J1148+1840}\\ 
  	              &                                                              &1&\cotwo &(224.70, 226.58) &225.6 &No & &\\ 
                  &                                                              &2& cont. &(236.65, 238.53) &237.6 &No & &\\ 
                  &                                                              &3& cont. &(238.85, 240.72) &239.8 &No & &\\
\hline
\multirow{4}{*}{2019.1.00036.S}&\multirow{4}{*}{{\tt uid\_A002\_Xea90c0\_X48c2}} &0& cont. &(222.70, 224.58) &223.6 &No &\multirow{4}{*}{J1058+0133}&\multirow{4}{*}{J1148+1840}\\ 
  	              &                                                              &1&\cotwo &(224.70, 226.58) &225.6 &No & &\\ 
                  &                                                              &2& cont. &(236.65, 238.53) &237.6 &No & &\\ 
                  &                                                              &3& cont. &(238.85, 240.72) &239.8 &No & &\\
\hline  
\end{tabular} \\
\tablecomments{(1) Project code. (2) Observational track ID. (3) SPW ID. (4) Targeted line of each SPW. (5) Frequency range of each SPW. (6) Central frequency of each SPW. (7) Whether the SPW is flagged manually. (8)--(9) Quasars used to do bandpass and phase calibration.} 
\end{table*}

\section{ALMA Observation and Reduction}\label{sec:alma_data}

\subsection{Observations \& Data Reduction}\label{sec:reduction}

NGC~3862 was observed with ALMA during Cycle~7 (PI: Dieu Nguyen; PID: 2019.1.00036.S) using the 12-m array in the C-5 configuration of 41 antennae. The primary beam has a full width at half maximum (FWHM) of $\approx$25\arcsec\ at Band~6. A summary of the observing tracks is provided in Table~\ref{tab:Observing Tracks}.

Each execution block was observed in Band~6 using four spectral windows (SPWs) in frequency-division mode (FDM). Each SPW spans 1.875~GHz with 1920 channels, yielding a native channel width of 976.562~kHz ($\approx$1.3~\kms). One SPW was tuned to the redshifted \cotwo\ line ($\nu_{\rm rest} = 230.538$~GHz), while the remaining three SPWs were dedicated to continuum measurement. The full spectral setup is listed in Table~\ref{tab:Spectral windows}.

All data were calibrated using the Common Astronomy Software Applications package \citep[\textsc{casa};][]{Hunter2023, Masters2020, CASATeam2022}\footnote{\url{https://casa.nrao.edu/}}, version~6.1.1.15, following the standard ALMA Science Pipeline\footnote{\url{https://almascience.eso.org/processing/science-pipeline}}. The pipeline performs standard interferometric calibration steps, including system temperature and water vapor radiometer corrections, bandpass calibration, gain calibration, and flux-density scale setting. The resulting calibrated measurement set (MS) is summarized in Tables~\ref{tab:Observing Tracks} and \ref{tab:Spectral windows}.

\begin{table}
\centering
	\caption{Parameters of the 1.3~mm continuum image and source.}
	\label{tab:continuum}
    \vspace{-3mm}
	\begin{tabular}{lc} 
	\hline\hline
Image property                & Value                               \\
    \hline
Image size (pixel$^2$)        & $256\times256$                      \\
Image size (arcsec$^2$)       & $25.6\times25.6$                    \\
Image size (kpc$^2$)          & $12.34\times12.34$                  \\
Pixel scale (arcsec per pixel)& 0.1                                 \\
Pixel scale (pc per pixel)    & 48.2                                \\
Sensitivity (mJy per beam)    & 0.145                               \\
Synthesised beam (arcsec$^2$) & $0.337\times0.284$                  \\
Synthesised beam (pc$^2$)     & $162\times137$                      \\
	\hline
Source property               & Value                               \\
	\hline
Right ascension               & $11^{\rm h}45^{\rm m}05\fs009$      \\
Declination                   & ${+}19\degr36\arcmin22\farcs741$    \\
Integrated flux (mJy)         & $73.57\pm 1.21$                     \\
Deconvolved size (arcsec$^2$) & $(0.36\pm 0.01)\times(0.30\pm 0.01)$\\
Deconvolved size (pc$^2$)     & $(172.6\pm 1.8)\times(144.0\pm1.3)$ \\
	\hline
	\end{tabular}
\end{table}

\begin{table}
\hspace{-12mm}
    \caption{The combined \cotwo\ data cube properties}
    \label{tab:co}
    \begin{tabular}{lr}
    \hline\hline
CO image property                     & Value                \\ \hline
Spatial extent (pixel$^2$)            & 256 $\times$ 256     \\ 
Spatial extent (arcsec$^2$)           & 25.6 $\times$ 25.6   \\ 
Spatial extent (kpc$^2$)              & 12.34 $\times$ 12.34 \\ 
Pixel scale (arcsec per pixel)        & 0.1                  \\ 
Pixel scale (pc per pixel)            & 48.2                 \\ 
Velocity range (km s$^{-1}$)          & 6200 -- 6500         \\ 
Channel width (km s$^{-1}$)           & 10                   \\ 
Number of velocity/frequency channels & 46                   \\
Mean synthesised beam (arcsec$^2$)    & 0.348 $\times$ 0.293 \\ 
Mean synthesised beam (pc$^2$)        & $168\times141$       \\ 
Sensitivity (mJy per beam per 10 \kms)& 0.41                 \\
    \hline
    \end{tabular}
\tablecomments{The kinematic center of the \cotwo\ CND is assumed to coincide with the peak of the 1.3~mm continuum emission; see Table~\ref{tab:continuum} for the continuum source parameters.} 
\end{table}

\begin{figure*}[!th]
    \centering
    \includegraphics[width=0.98\textwidth]{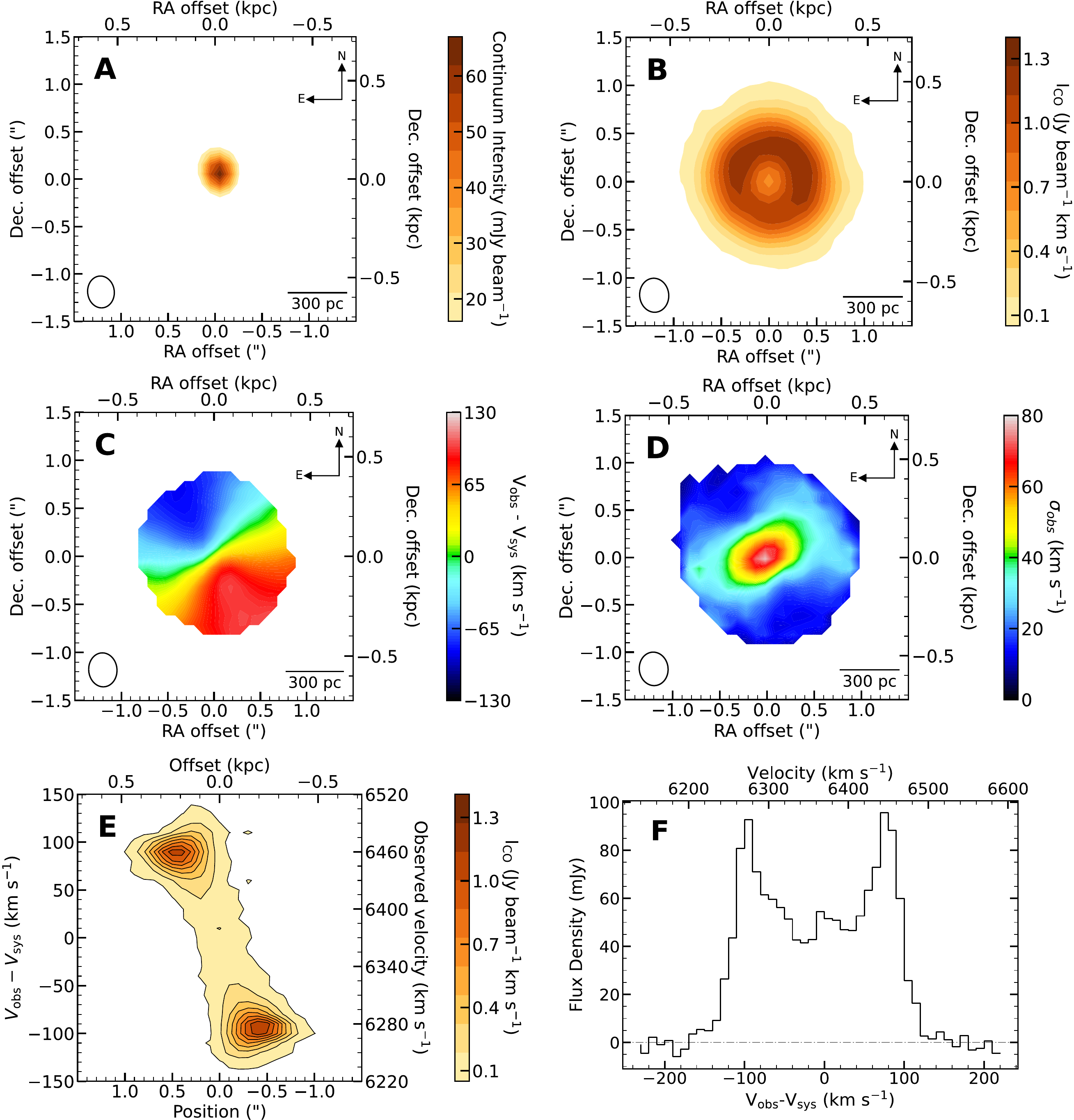}
    \caption{ALMA Band~6 observations of the \cotwo-molecular gas in NGC~3862. {\it Panel~A:} 1.3~mm continuum emission. Contours begin at 110$\sigma$ and increase in steps of 35$\sigma$, where $1\sigma = 0.145~\rm{mJy~beam}^{-1}$. {\it Panel~B:} \cotwo\ integrated intensity (moment-0) map. {\it Panel~C:} Intensity-weighted mean LOS velocity (moment-1) map, referenced to the systemic velocity $v_{\rm sys} = 6370$~\kms\ (see Section~\ref{sec:momentmaps}). {\it Panel~D:} Intensity-weighted LOS velocity dispersion (moment-2) map. In Panels~A--D, the synthesized beam ($\theta = 0\farcs 348 \times 0\farcs 293$, PA = 0$^\circ$) is shown as a filled black ellipse in the lower-left corner, and the map center is set to the best-fit kinematic center (R.A., decl.) $ = $ ($11^{\rm h}45^{\rm m}05\fs009$, ${+}19\degr36\arcmin22\farcs741$). {\it Panel~E:} Major-axis PVD extracted along PA~$= 209\arcdeg$ through the kinematic center, with velocities referenced to $v_{\rm sys}$.  {\it Panel~F:} \cotwo\ integrated spectrum extracted within a $3.0\arcsec \times 3.0\arcsec$ ($1.45~{\rm kpc} \times 1.45~{\rm kpc}$) aperture centered on the kinematic center. The horizontal dot-dashed line marks zero flux density. The double-horned profile is characteristic of emission from a rotating disk.}
    \label{fig:moment-maps}
\end{figure*}

\subsection{Continuum Imaging}\label{sec:cont}

The 1.3~mm continuum image was produced using the \textsc{casa} task \texttt{tclean} in multifrequency synthesis mode \citep{Rau11}. The continuum was constructed by combining all three continuum SPWs with the line-free channels of the \cotwo\ SPW from the MS. Briggs weighting with a robust parameter of 0.5 was applied to balance angular resolution and point-source sensitivity. The resulting synthesized beam is $\theta_{\rm cont} = 0\farcs 337 \times 0\farcs 284$ at a position angle of ${\rm PA} = 7\fdg 0$, with a root-mean-squared (RMS) noise of $\sigma_{\rm cont} = 145~\rm{mJy~beam}^{-1}$.

The continuum map reveals a single, spatially resolved source with a deconvolved size of $0\farcs 36 \times 0\farcs 30$, as confirmed by a two-dimensional (2D) Gaussian fit using the \textsc{casa} task \texttt{imfit}. The source is coincident with the galaxy kinematic center (i.e., the best-fit SMBH position; Section~\ref{sec:results}) and has an integrated flux density of $S_{\rm 1.3\,mm} = 73.57 \pm 1.21$~mJy, consistent with the nominal $\approx$10\% ALMA absolute flux calibration uncertainty. The continuum image is shown in Figure~\ref{fig:moment-maps} (Panel~A), and the imaging parameters and source properties are summarized in Table~\ref{tab:continuum}.

\subsection{Creating the \cotwo\ Data Cube}\label{sec:line}

The 1.3~mm continuum emission is detected at the nucleus of NGC~3862 and is spatially resolved in the MS (Section~\ref{sec:reduction}; Table~\ref{tab:Spectral windows}). The continuum was modeled with a first-order polynomial in the $uv$-plane using all line-free channels across the four SPWs (Table~\ref{tab:Spectral windows}), and subsequently subtracted from the calibrated visibilities using the \textsc{casa} task \texttt{uvcontsub} \citep[e.g.,][]{Davis_2020, Nguyen_2020}.

The continuum-subtracted, three-dimensional~(3D)~\cotwo\ spectral-line cube (R.A., Decl., velocity) was produced using the \textsc{casa} task \texttt{tclean} with a H\"{o}gbom deconvolver \citep{Hogbom74}, Briggs weighting with a robust parameter of 0.5, and a channel width of 10~\kms. This channel width is well-suited for SMBH dynamical modeling \citep{Davis_2014, Nguyen_2021, Ngo_2025_7052}. The spatial grid consists of $256 \times 256$ pixels at a scale of $0\farcs 10$~pixel$^{-1}$, providing adequate Nyquist sampling of the synthesized beam. The velocity axis is referenced to the \cotwo\ rest frequency of $\nu_{\rm rest} = 230.538$~GHz.

During interactive deconvolution, a \texttt{clean} mask was applied to regions of significant emission to suppress sidelobe artifacts and improve image fidelity. Cleaning proceeded to a threshold of $1.5\,\sigma_{\rm RMS}$, where $\sigma_{\rm RMS}$ is measured from line-free channels in the dirty cube. The final cleaned cube has a synthesized beam identical to that of the 1.3~mm continuum image (Section~\ref{sec:cont}), and its properties are summarized in Table~\ref{tab:co}.

\subsection{\cotwo\ Emission Moment Maps}\label{sec:momentmaps}

The \cotwo\ emission spans a velocity range of 6200--6500~\kms\ relative to a systemic velocity of $v_{\rm sys} \approx 6370$~\kms. Moment maps were derived from the cleaned data cube using the masked-moment method \citep{Dame11}. The zeroth (integrated intensity, Panel~B), first (intensity-weighted mean line-of-sight (LOS velocity, Panel~B), and second (intensity-weighted LOS velocity dispersion, Panel~D) moment maps are shown in Figure~\ref{fig:moment-maps}. We note that the moment maps are not beam-deconvolved.

The mask was constructed as follows. A copy of the cleaned cube was smoothed spatially with a Gaussian kernel of width $\sigma = 1.5\,\theta_{\rm FWHM}$ and spectrally with a Hanning window of four times the native channel width \citep{Smith_2021, Dominiak_2024_4751}. A binary mask was then defined by applying a $0.5\,\sigma_{\rm RMS}$ threshold to the unsmoothed cube, equivalent to $0.6\,\sigma_{\rm RMS}$ in the smoothed cube. This approach effectively suppresses noise while preserving the majority of the line flux. The resulting mask includes 4348 pixels above the threshold, which were used to extract the moment maps from the unsmoothed cube.

The zeroth-moment map reveals that the \cotwo\ CND has a nearly circular morphology with an apparent diameter of $\approx$2\arcsec\ (964~pc). The integrated intensity peaks in an annular structure at a radius of $\approx$0.3\arcsec\ from the disk center, with a central depression coincident with the AGN position, suggesting that the molecular gas is distributed in a ring-like geometry with reduced emission at the nucleus. The outer contours exhibit mild asymmetries, probably caused by a slightly disturbed or warped disk. The spatial distribution of the \cotwo\ emission is broadly consistent with the circumnuclear dust disk identified in \hst\ imaging (Figure~\ref{fig:HST-overlaid}).

The total molecular gas mass was estimated from the integrated \cotwo\ flux density using the CO-to-H$_2$ conversion factor $X_{\rm CO} = 2 \times 10^{20}$~cm$^{-2}$~(K~km~s$^{-1}$)$^{-1}$ \citep[$\alpha_{\rm CO} = 4.3$~\Msun~(K~\kms~pc$^2$)$^{-1}$;][]{Bolatto2013}:
\begin{align}
M_{\rm gas} = & 1.05 \times 10^4 
\left( \frac{X_{\rm CO}}{2 \times 10^{20}~{\rm cm^{-2}~(K~km~s^{-1})^{-1}}} 
\right) \\ \nonumber
& \times \left( \frac{1}{1+z} \right) 
\left( \frac{S_{\rm CO}\Delta v}{{\rm Jy~km~s}^{-1}} \right) 
\left( \frac{D_L}{{\rm Mpc}} \right)^2 \, {\rm M}_{\odot},
\label{eq:mgas}
\end{align}
where $S_{\rm CO}\,\Delta v = 12.97$~Jy~\kms\ is the integrated \cotwo\ flux density measured from our data cube. We adopt $z = 0.0217$ \citep{Casadei2024} and $D_L =D = 99.4$~Mpc for NGC~3862 \citep{Ma_2014}, and assume a \cotwo-to-\coone\ line flux ratio of unity \citep{Smith_2021}. Under these assumptions, the total molecular gas mass is $M_{\rm gas} \approx 1.32 \times 10^9$~\Msun.

The first-moment map (Panel~C) reveals a regularly rotating velocity field with a smooth transition from receding (northwestern) to approaching (southeastern) velocities, reaching $\pm$130~\kms\ relative to $v_{\rm sys}$. The well-ordered, symmetric velocity contours are consistent with a geometrically thin, nearly coplanar rotating disk with no evidence of significant warping across the observed extent of the CND.

The second-moment map (Panel~D) shows that the intensity-weighted LOS velocity dispersion is low across most of the disk ($\sigma_{\rm LOS} \lesssim20$~\kms), rising sharply to $\sigma_{\rm LOS} \approx 60$--$80$~\kms\ in the central region. The elevated dispersion is spatially elongated along the kinematic major axis, a morphology characteristic of beam smearing \citep{Davis17}, in which emission from regions of substantially different LOS velocities is artificially blended within the synthesized beam \citep{Barth16b}.  The magnitude of this central dispersion enhancement is consistent with the steep velocity gradient expected within the SMBH SOI, and does not reflect intrinsic turbulence in the molecular gas.

\subsection{Integrated Spectrum \& Position--Velocity Diagram}\label{sec:pvd}

Panel~E shows the position--velocity diagram (PVD) of the \cotwo\ CND, extracted along the kinematic major axis at PA = 209\arcdeg\ using a 4-pixel-wide ($0\farcs4$) pseudo-slit. Pixels were selected above $0.5\,\sigma_{\rm RMS}$ in the unsmoothed data cube after applying a spatial Gaussian filter with a FWHM equal to the synthesized beam. The PVD reveals two distinct emission concentrations symmetrically offset from the disk center, at $\approx$$+0\farcs8$/$+100$~\kms\ and $\approx$$-0\farcs7$/$-100$~\kms, with a central gap in emission coincident with the AGN position. This bilobular morphology is consistent with the ring-like gas distribution identified in the zeroth-moment map (Panel~B), and indicates that the molecular gas does not peak in the innermost region surrounding the SMBH. The emission extends to $\approx$$\pm 1\farcs2$ ($\approx$$\pm 0.6$~kpc) along the major axis, providing leverage on the stellar \ml\ and thereby tightening the constraints on \Mbh\ in the dynamical models (Section~\ref{sec:results}).

Panel~F shows the integrated \cotwo\ spectrum of NGC~3862, extracted within a $3.0\arcsec \times 3.0\arcsec$ ($1.45 \times 1.45$~kpc) aperture enclosing all detected line emission. The spectrum displays a characteristic double-horned profile, with peaks at $\approx$$\pm100$~\kms\ relative to $v_{\rm sys}$, spanning a total velocity extent of $\approx$280~\kms, broadly consistent with a regularly rotating, spatially resolved CND.

\section{Galaxy Stellar-Mass Model}\label{sec:stellar-mass}

\subsection{HST Imaging and PSF Model}\label{sec:hst-psf}

NGC~3862 was observed with {\it HST}/WFC3/UVIS in the F814W and F475W filters under Program ID~14159 (PI: E.~T.~Meyer) on 2015 November~16. Each filter was observed with $N_{\rm exp} = 3$ individual exposures, yielding total exposure times of 1098~s per filter. The calibrated, flat-fielded images were retrieved from the Mikulski Archive for Space Telescopes (MAST\footnote{\url{https://mast.stsci.edu/}}). Given that the F475W band is more heavily affected by dust extinction, we adopt the F814W image as the basis for the stellar mass model in our fiducial analysis.

Point spread functions (PSFs) for both filters were generated using the \textsc{TinyTim}\footnote{\url{https://github.com/spacetelescope/tinytim/releases/tag/7.5}} package \citep{Krist95, Krist11}. Individual model PSFs were computed for each exposure based on the instrument configuration, detector position, and dither offset, and each was convolved with the appropriate charge-diffusion kernel to account for CCD electron diffusion effects. The per-exposure PSFs of each filter were then combined and resampled to a final pixel scale of $0\farcs04$ using \textsc{AstroDrizzle}\footnote{\url{https://www.stsci.edu/scientific-community/software/drizzlepac}} within the \texttt{DrizzlePac} package \citep{Avila12}, matching the dither pattern of the observations.

\begin{table}
\caption{MGE PSF model of \hst/WFC3 F814W image \label{tab:mge_psf}}
\begin{tabular}{cccc}
\hline\hline
$j$&(Light fraction)$_j$&$\sigma_j~({\rm arcsec})$&$q_j=b_j/a_j$ \\
(1) & (2) & (3) & (4) \\
\hline
1 & 0.563 & 0.021 & 1.000\\
2 & 0.263 & 0.059 & 1.000\\
3 & 0.056 & 0.195 & 1.000\\
4 & 0.075 & 0.406 & 1.000\\
5 & 0.043 & 1.005 & 1.000\\
\hline
\end{tabular}
\tablecomments{(1)~Gaussian component index; (2)~light fraction; (3)~dispersion along the major axis; (4)~axial ratio of Gaussian components.}
\end{table}

\begin{figure}
    \centering\includegraphics[width=0.98\columnwidth]{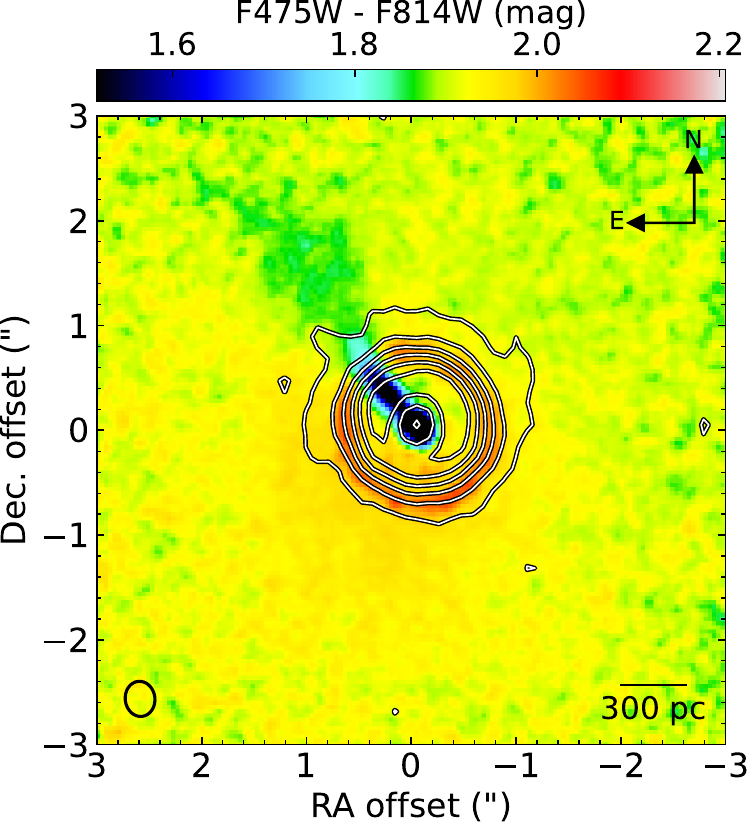}
    \caption{HST/WFC3 F475W$-$F814W color map of the nuclear region of NGC 3862, with ALMA \cotwo\ integrated intensity contours overlaid in white; contour levels and synthesized beam as in Figure~\ref{fig:HST-overlaid}. The stellar color is uniform at $\approx$1.9--2 mag away from the central blue jet feature (NE--SW) and the dust lane within $r\lesssim1\arcsec$, consistent with a single, old ($\gtrsim$10 Gyr) stellar population dominating the nuclear light.}
    \label{fig:F475W-F814W}
\end{figure}

\subsection{Dust Extinction Correction for the F814W image}\label{sec:dust}

As shown in Figure~\ref{fig:HST-overlaid}, the circumnuclear dust disk produces a clearly visible absorption feature against the stellar continuum within $r \lesssim 1\arcsec$ of the nucleus, coincident with the \cotwo\ CND. If left uncorrected, this extinction would suppress the observed surface brightness in the radial range that provides the strongest dynamical constraint on \Mbh, affecting the innermost MGE Gaussian components and the resulting \Mbh. We therefore correct the F814W image for dust extinction before fitting the MGE (Section~\ref{sec:mge}), following the color-excess method \citep{Cappellari_2002_dust, Sabine_2019}, which assumes that the dust acts as a foreground screen and reddens the observed stellar light without changing its intrinsic color.

We first constructed an F475W$-$F814W color map from the sky-subtracted and masked F475W and F814W images. The sky background was estimated as the median pixel value within several $20 \times 20$~pixel$^2$ apertures placed in source-free regions at projected radii $>$50\arcsec\ from the galaxy center, and was then subtracted uniformly from each image. We also created photometric masks for both images to exclude foreground stars and cosmetic defects (bad and hot pixels). These sources were identified using a point-source catalog generated with \textsc{SExtractor}\footnote{\url{https://www.astromatic.net/software/sextractor/}} \citep{Bertin96} and masked accordingly.

\begin{figure}
    \centering
    \includegraphics[width=0.98\columnwidth]{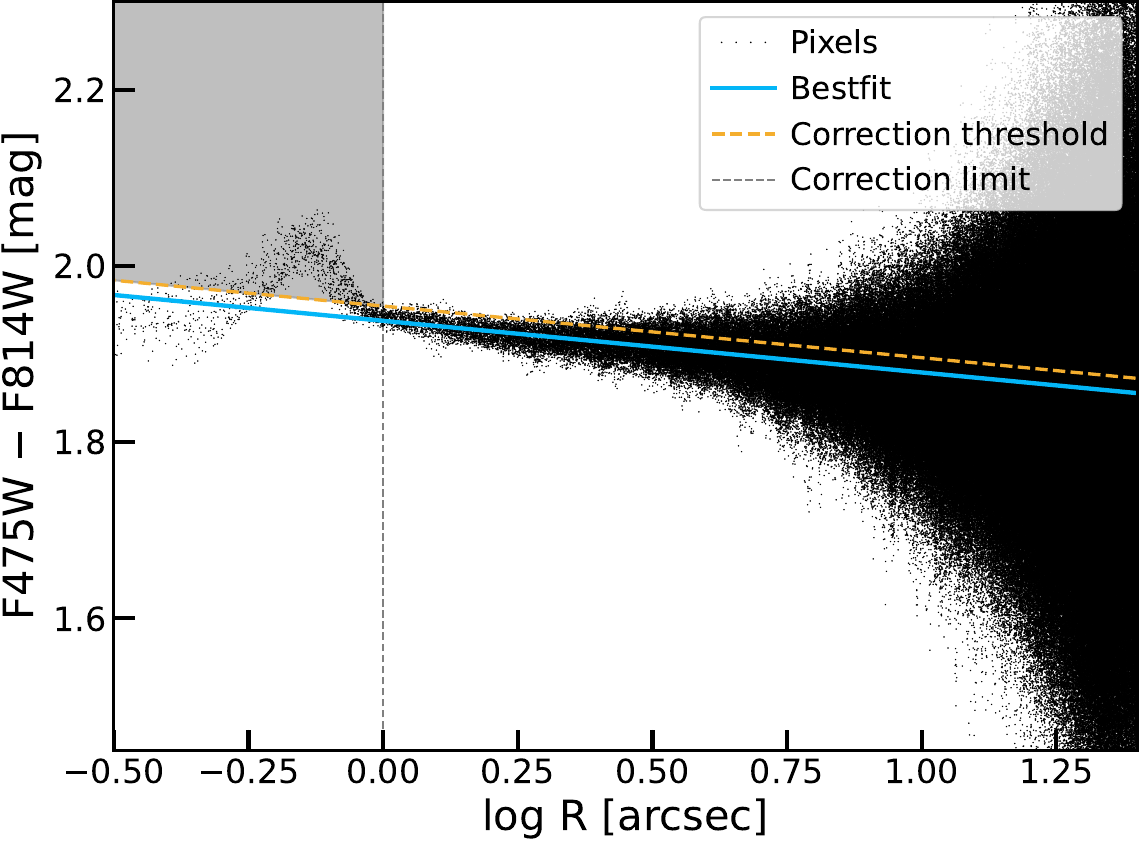}
    \caption{HST/WFC3 F475W$-$F814W pixel color across the nucleus of NGC~3862 versus $\log R$, where $R$ is the elliptical (semi-major-axis) radius, with the synchrotron jet and central AGN masked out. The blue line is the best-fit intrinsic color relation; the orange dashed line is the correction threshold, offset by $+1\sigma$ ($E = 0.017$~mag). Pixels above the threshold and within $r < 1\arcsec$ (gray shaded region) were corrected for dust extinction.}
    \label{fig:color_scatter}
\end{figure}

\begin{figure}
    \centering
    \includegraphics[width=0.98\columnwidth]{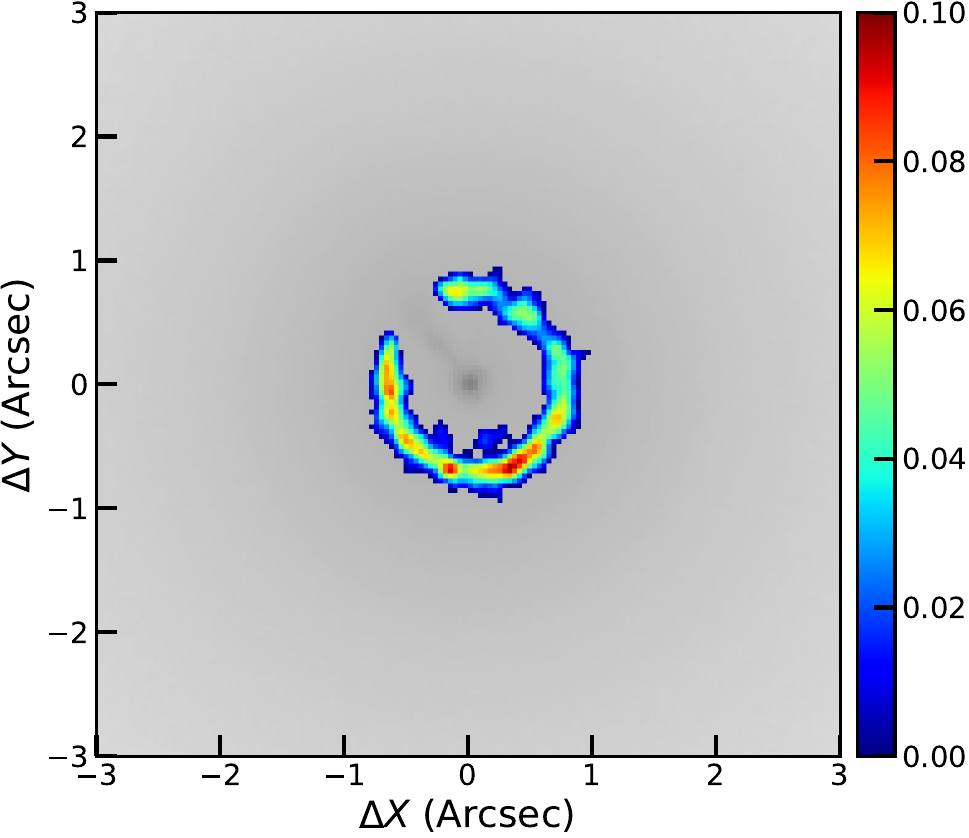}
    \caption{Central region of the dust-corrected HST/WFC3 F814W image of NGC~3862. The color overlay marks the corrected pixels, shaded by the fractional flux increase they received (0.1 corresponds to a 10\% brightening); the correction traces a ring coincident with the CND.}
    \label{fig:correction_degree}
\end{figure}

One step before constructing the color map, we astrometrically aligned the sky-subtracted and masked images and cross-convolved them with each other's PSFs---the F475W image with the F814W \textsc{TinyTim} PSF and vice versa---to minimize artificial color gradients arising from differences in PSF width between the two filters. The resulting F475W$-$F814W color map, shown in Figure~\ref{fig:F475W-F814W}, reveals a uniform stellar color of $\approx$2~mag across the central $6\arcsec \times 6\arcsec$ region, with two localized exceptions: a blue feature coincident with the central AGN and synchrotron jet, and a redder region within $r\lesssim1\arcsec$ tracing the circumnuclear dust lane. Away from these regions, this uniformity is consistent with a single, old stellar population ($\gtrsim$10~Gyr).

Second, we measured the F475W$-$F814W color of every pixel in the central region as a function of its elliptical radius (semi-major axis, in logarithmic scale), after masking the synchrotron jet and the central AGN-contaminated pixels. The resulting color--$\log r$ distribution is shown in Figure~\ref{fig:color_scatter}.

\begin{figure*}
\centering
\includegraphics[width=0.95\textwidth]{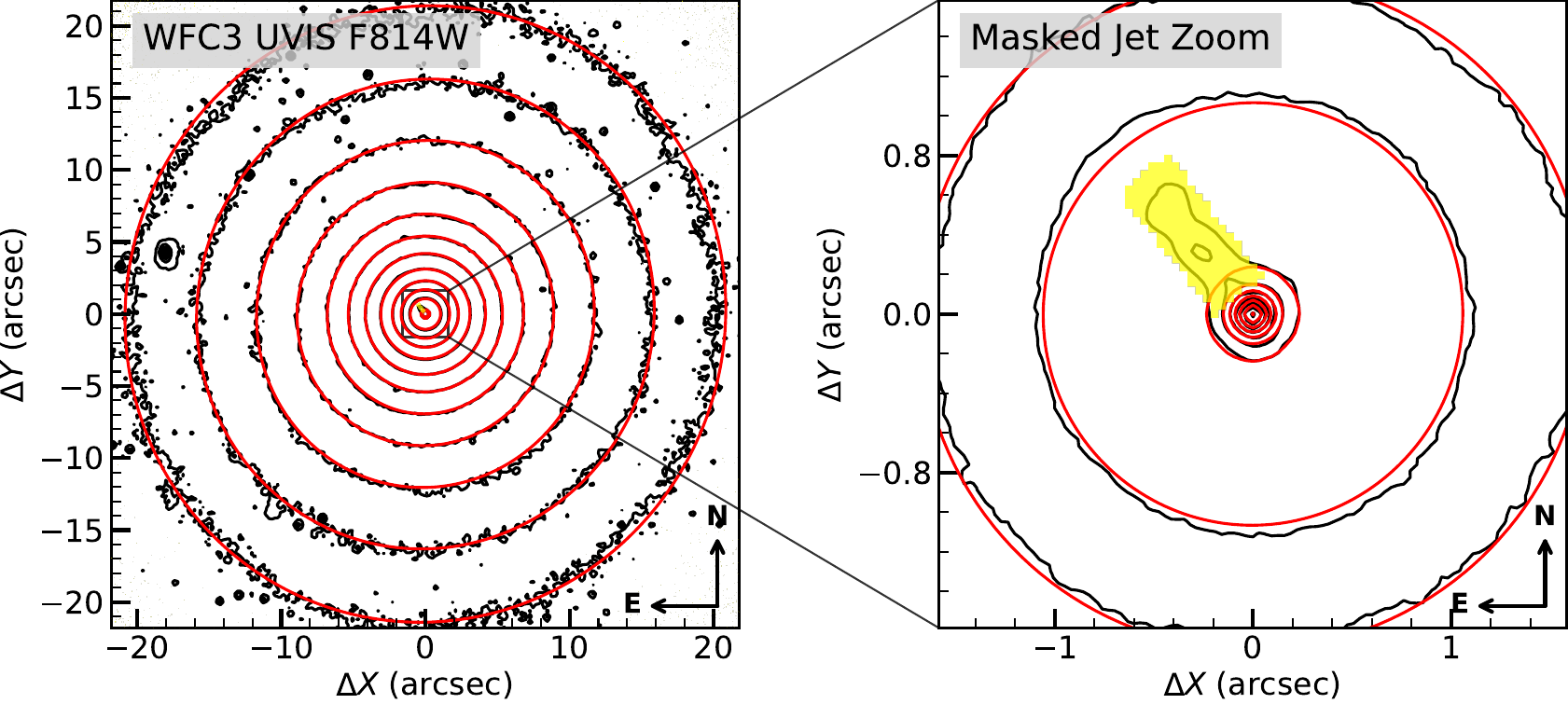}
\caption{MGE surface brightness modeling of NGC~3862 from the sky-subtracted, masked, and dust-corrected F814W image with black contours show the data, and red contours show the MGE model. {\it Left:} Large-scale view over a $44\arcsec \times 44\arcsec$ field, showing the global MGE fit to the stellar light distribution.  {\it Right:} Zoom-in of the central $3.2\arcsec \times 3.2\arcsec$ region. Here, contours trace the stellar continuum intensity in steps of 1 mag arcsec$^{-2}$. The yellow region marks pixels affected by the jet, which were excluded from the MGE fit.}
\label{fig:mge-fit}
\end{figure*}

To separate the intrinsic stellar color gradient from reddening by dust, we assumed that the intrinsic color varies linearly with $\log r$ \citep{Cappellari_2002_dust, Sabine_2019}. We fit this relation (blue line in Figure~\ref{fig:color_scatter}) with a robust linear regression to pixels in the dust-free region ($1\arcsec \le r \le 5\arcsec$, outside the jet cone), where dust has a negligible influence. The color excess of each pixel, $E({\rm F475W}-{\rm F814W})$, is then its observed color minus this intrinsic relation.

We adopted a correction threshold of $E({\rm F475W}-{\rm F814W}) = 0.017$~mag, equal to the $1\sigma$ scatter of the fit (yellow dashed line in Figure~\ref{fig:color_scatter}). Within $r < 1\arcsec$, pixels redder than this threshold were identified as dust-affected and corrected in the F814W band through $A_I = k\,E({\rm F475W}-{\rm F814W})$, with $k = A_{814}/(A_{475}-A_{814}) = 0.98$ from the CCM89 Galactic extinction law \citep[$R_V = 3.1$;][]{Cardelli1989}. Pixels at $r > 1\arcsec$, and those bluer than the intrinsic relation, were left unchanged; the central blue excess traces the synchrotron jet rather than dust absorption. The correction is confined to $r < 1\arcsec$ and reaches at most $\approx$10\% of the F814W flux in the most obscured pixels, tracing a ring that coincides with the CND (Figure~\ref{fig:correction_degree}). This dust-corrected F814W image is used as the input to the fiducial Multi-Gaussian Expansion \citep[MGE\footnote{v6.0.4: \url{https://pypi.org/project/mgefit/}};][]{Emsellem94} fit (Section~\ref{sec:mge}), and the systematic impact of the correction on \Mbh\ and \ml$_{\rm F814W}$ is assessed in Section~\ref{sec:uncertainties}.

\begin{table}[h]
\caption{\hst/WFC3 UVIS F814W MGE models}  
\setlength{\tabcolsep}{2pt}
\begin{tabular}{ccccc} 
\hline\hline
$j$ &$\log I_j$(\Lsun\ ${\rm pc^{-2}}$)&$\log\sigma_j$(arcsec)&$q'_j=b_j/a_j$& Comp.\\
(1) & (2) & (3) & (4) & (5) \\ 
\hline
1 & 5.939 & $-$1.735 & 0.997 & AGN     \\
2 & 3.732 & $-$1.005 & 1.000 & stars \\
3 & 3.701 &    0.003 & 0.995 & stars \\
4 & 3.222 &    0.374 & 0.995 & stars \\
5 & 2.619 &    0.703 & 0.967 & stars \\
6 & 2.133 &    1.175 & 0.969 & stars \\
\hline
\end{tabular}
\tablecomments{(1)~Gaussian component index; (2)~molecular gas mass surface density; (3)~Gaussian dispersion along the major axis; (4)~axial ratio. Here, the position angles of all Gaussian components are fixed to $\rm{PA} = 209\degr$ during the MGE fitting, motivated by the observed alignment between the \cotwo\ CND and the large-scale stellar distribution of NGC~3862.}
\label{table:mge-table}
\end{table}

\subsection{Stellar Photometric Models}\label{sec:mge}

The nucleus of NGC~3862 is contaminated by AGN continuum and jet emission \citep{Meyer2015}, the dominant source of systematic uncertainty in our \Mbh\ measurement. We therefore masked the jet-affected pixels by hand, as shown in the right panel of Figure~\ref{fig:mge-fit}, before fitting the stellar light with an MGE.

We modeled the stellar light distribution of NGC~3862 following \citet{Cappellari2002}, in two steps. First, we decomposed the HST/WFC3 UVIS F814W \textsc{TinyTim} PSF (Section~\ref{sec:hst-psf}) into an MGE with \texttt{mge\_fit\_sectors}; the resulting Gaussians are listed in Table~\ref{tab:mge_psf}. Second, we supplied this PSF MGE to \texttt{mge\_fit\_sectors\_regularized} and fit the sky-subtracted, masked, and dust-corrected F814W image with a set of 2D Gaussians, convolved with the PSF during the fit. The routine fits the surface brightness along multiple photometric sectors, with regularization applied to suppress unphysical oscillations between adjacent Gaussians.

The position angle of all Gaussian components was fixed at ${\rm PA} = 209\arcdeg$ during the MGE fit, matching the orientation of both the large-scale stellar isophotes and the \cotwo\ CND (Section~\ref{sec:pvd}). To check that this does not bias the stellar mass model, we repeated the fit with the PA free using \texttt{mge\_fit\_sectors\_twist} \citep{Cappellari2002}; the best-fit PAs scatter within $210\arcdeg \pm 8\arcdeg$, consistent with the fixed value, so the assumption has a negligible effect on \Mbh.

For the photometric calibration of the F814W image, we adopted a zero-point magnitude $m_{\rm ZP} = 24.686$~mag \citep{Nguyen_2019}, a foreground Galactic extinction $A_{\rm F814W} = 0.036$~mag from the NASA/IPAC Extragalactic Database (NED\footnote{\url{https://ned.ipac.caltech.edu}}), and an $I$-band solar absolute magnitude $M_{\odot,I} = 4.12$~mag \citep{Willmer18}, all in the Vega system. The resulting F814W MGE model is listed in Table~\ref{table:mge-table}. Because the central pixels are likely contaminated by AGN light, we treat the smallest Gaussian as the AGN contribution and exclude it from the stellar MGE. We then converted the deconvolved stellar MGE into a mass distribution by scaling it with a spatially constant $M/L_{\rm F814W}$. Dark matter is neglected, as the \cotwo\ CND extends only to $1\arcsec$ from the nucleus, well within the region where stars dominate the gravitational potential \citep{Cappellari13a, Nguyen_2026_highz}. The total gravitational mass model of NGC~3862 therefore comprises three components: a central point mass for the SMBH (\Mbh), the stellar mass distribution, and the molecular gas mass distribution.

\section{Dynamical Modeling}\label{sec:dynamical-models}

\subsection{\kinms\ Tool} \label{sec:kinms}

We modeled the \cotwo\ CND kinematics of NGC~3862 to constrain its \Mbh\ using the \textsc{Python} implementation of the \kinms\ package \citep{Davis_2013, Davis_2014}, which has been widely applied in the mm-Wave Interferometric Survey of Dark Object Masses \citep[WISDOM; e.g.,][]{Davis17, Onishi17, Ruffa23} and the Measuring Black Holes in Milky Way-mass Galaxies \citep[MBHBM$_\star$;][]{Nguyen_2020, Nguyen_2022} projects.

\kinms\ forward-models the observed data cube by simulating the 3D gas distribution and kinematics, while self-consistently incorporating observational effects including beam smearing, spatial and spectral binning, and LOS projection. The simulated cube is compared directly to the observed cube to infer the best-fit parameters and their uncertainties via a Markov chain Monte Carlo (MCMC) $\chi^2$ minimization routine with Bayesian priors (Section~\ref{sec:bayesian-infer}).

We assume that the \cotwo\ gas follows circular orbits in the combined gravitational potential of the SMBH, stars, and molecular gas. The circular velocity curve was computed as a function of radius at each iteration using the \texttt{mge\_vcirc} routine from the Jeans Anisotropic Modeling \citep[\textsc{jam}\footnote{v7.2.4: \url{https://pypi.org/project/jampy/}};][]{Cappellari08, Cappellari2020} framework, accounting for both the stellar-mass (Table~\ref{table:mge-table}) and the molecular gas mass distribution, which was included via the \texttt{gasGrav} mechanism in \kinms, under the assumption that the dust and molecular gas are cospatially distributed.

\subsection{Gas Distribution} \label{sec:SkySampler}

Because \kinms\ fits the full 3D data cube, a parameterized description of the gas distribution is required. We represent the \cotwo\ spatial distribution using the \skysampler\footnote{\url{https://github.com/Mark-D-Smith/KinMS-skySampler}} package \citep{Smith19}, which samples gas cloud positions directly from the CLEAN components of the deconvolved data cube rather than assuming an analytic functional form for the gas morphology. This approach introduces a single additional free parameter, the integrated flux scaling factor ($f$), which rescales the model cube to match the observed total flux. Because CLEAN components represent only the deconvolved emission and exclude the residual map, their total flux is systematically lower than that of the cleaned cube; $f$ compensates for this deficit and ensures flux conservation within the modeled region.

The CLEAN components were sampled with $N = 10^6$ gas particles using the \texttt{sampleClouds} routine; this number of gas particles is sufficient to reproduce the observed \cotwo\ surface brightness distribution after beam convolution without introducing particle sampling noise. The particle positions were then deprojected from the sky plane to the intrinsic disk plane using the \texttt{transformClouds} routine, with the disk's position angle and inclination angle treated as free parameters.

\subsection{Gas Velocity Dispersion} \label{sec:dispersion}

The second-moment map (Panel~D of Figure~\ref{fig:moment-maps}) shows that the observed LOS velocity dispersion rises sharply from $\sigma_{\rm gas}=40$~\kms\ (at $r \lesssim 0\farcs4$) to $\sigma_{\rm gas}=80$~\kms\ at the center. This central enhancement is consistent with beam smearing rather than intrinsic turbulence: in regions unaffected by beam smearing, the velocity dispersion is uniformly $\lesssim$10~\kms, while the observed LOS rotation velocities reach $\pm$130~\kms. The ratio $\sigma_{\rm gas}/v_{\rm rot} \ll 1$ confirms that the molecular gas is dynamically cold and that circular orbits are a valid assumption.

We therefore include a non-zero, spatially uniform intrinsic turbulent velocity dispersion ($\sigma_{\rm gas}$) in the \kinms\ model as a free parameter, which acts as an isotropic broadening of the LOS velocity distribution of each gas particle, independent of the circular velocity field. The sensitivity of \Mbh\ and \ml$_{\rm F814W}$ to variations in $\sigma_{\rm gas}$ is assessed as part of our systematic uncertainty analysis in Section~\ref{sec:sigma}. We further model the \cotwo\ CND as a geometrically thin disk with zero intrinsic scale height, consistent with the dynamically cold nature of the gas. 

Our \kinms\ model has nine free parameters, summarized in Table~\ref{tab:mcmc-results}. The first two are the kinematic center coordinates ($x_{\rm cen}$, $y_{\rm cen}$), which specify the SMBH's position relative to the data cube phase center. We assume the kinematic center coincides with the peak of the 1.3~mm continuum emission (Section~\ref{sec:cont}), which is justified given that any offset between the two is expected to be much smaller than the synthesized beam. The third parameter is the systemic velocity of the gas disk ($v_{\rm sys}$), which sets the rest-frame velocity of the rotating disk. The fourth is the integrated flux scaling factor ($f$), which normalizes the model gas distribution to match the observed total flux. The next three parameters describe the CND geometry and intrinsic kinematics: the position angle (PA), inclination angle ($i$), and the intrinsic turbulent velocity dispersion of the molecular gas ($\sigma_{\rm gas}$). The final two parameters are the primary scientific quantities of interest: the SMBH mass (\Mbh) and the stellar mass-to-light ratio in the HST/WFC3 F814W band (\ml$_{\rm F814W}$).

\begin{figure*}
\centering
    \includegraphics[width=0.98\textwidth]{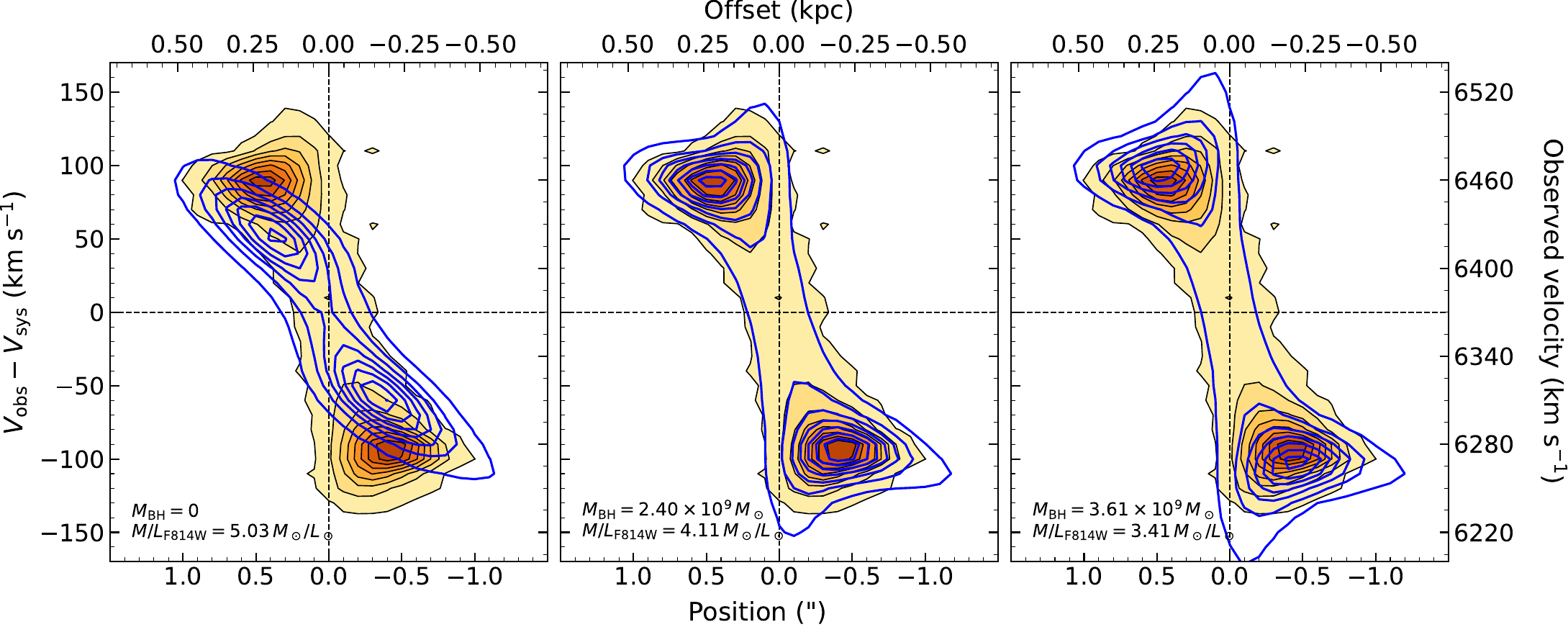}
    \caption{PVDs comparing the ALMA \cotwo\ observations (orange-filled contours) with \kinms\ models (blue contours). The PVDs are extracted along the galaxy’s major axis at a position angle of PA~$ = 209^\circ$ for three SMBH masses: no black hole (\textit{left }), the best-fitting SMBH (\textit{middle}), and an overly large SMBH (\textit{right}). The SMBH mass and \ml$_{\rm F814W}$ are indicated in the bottom-left of each panel. Black dashed lines mark the dynamical center, defined by the peak of the 1.3~mm continuum (Section~\ref{sec:cont}), with the intersection corresponding to the systemic velocity ($v_{\rm sys}=6370$~\kms) shown on the velocity axis. }
    \label{fig:compare-pvd}
\end{figure*}

\begin{figure*}[!th]
 \centering
    \includegraphics[width=0.98\textwidth]{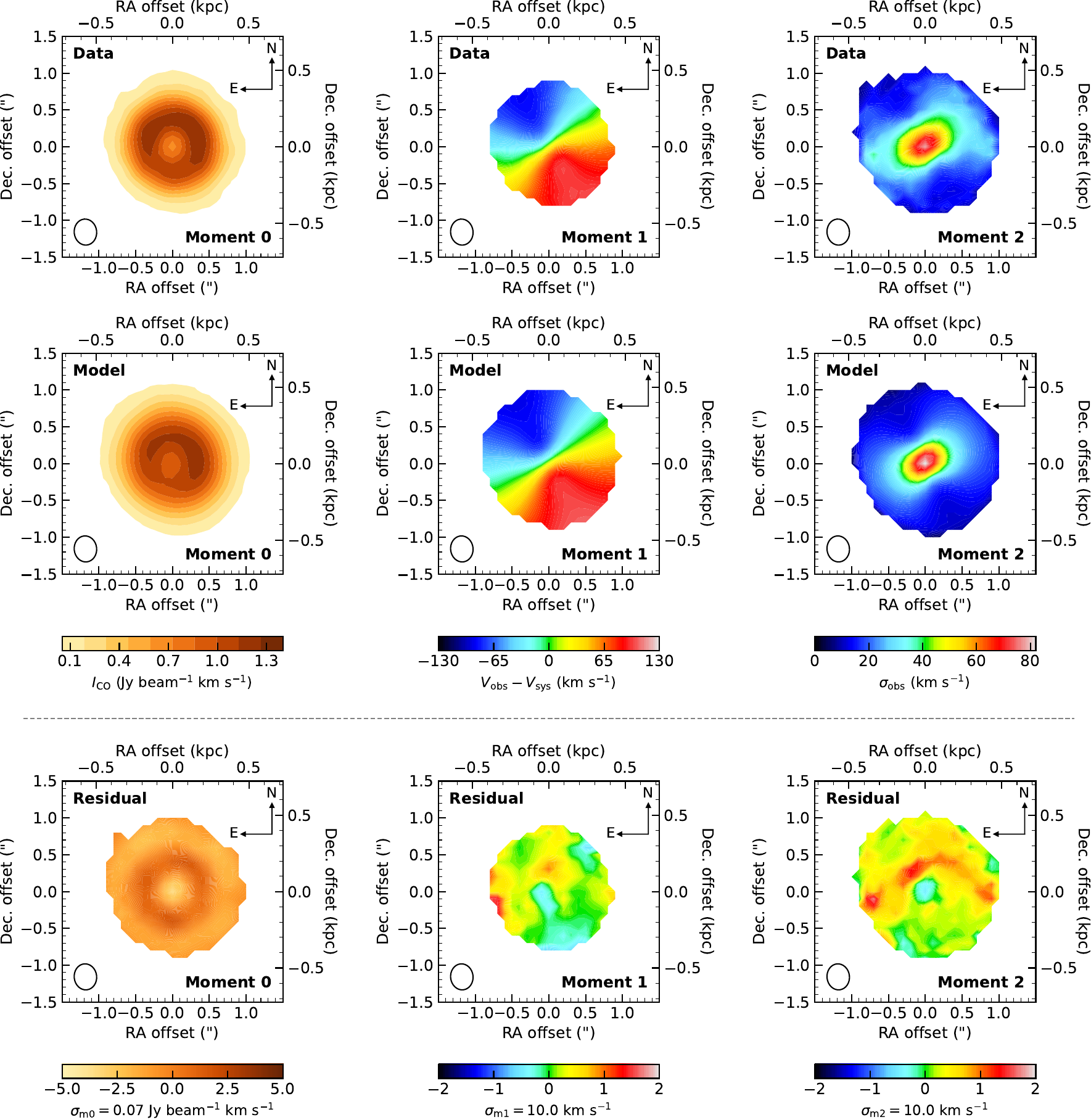}
    \caption{Comparison of the \cotwo\ moment maps from our ALMA observations (\textit{top-row panels}), the best-fitting \kinms\ model (\textit{middle-row panels}), and the normalized moment-residual maps, ({\tt data-model})/$\sigma$ (\textit{bottom-row panels}).} 
    \label{fig:momentmaps_compare}
\end{figure*}

\begin{figure}
    \centering
    \includegraphics[width=0.98\columnwidth]{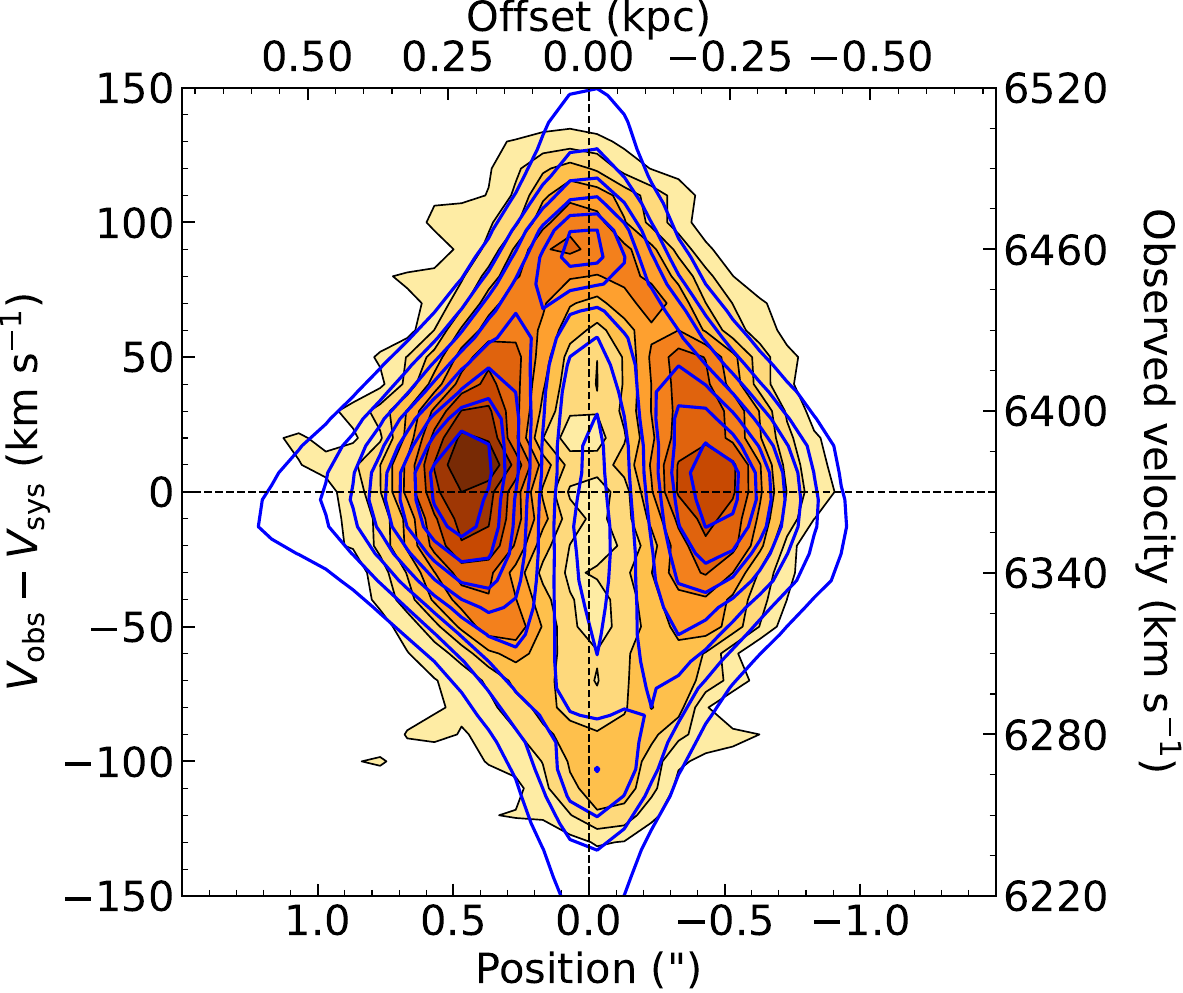}
    \caption{Minor-axis PVD of the \cotwo\ CND in NGC~3862, extracted along PA~$= 209^{\circ} + 90^{\circ}$ with velocities referenced to $v_{\rm sys} = 6370$~\kms. The best-fit \kinms\ model is overlaid in blue at the same contour levels as the data.}
    \label{fig:PVD_minoraxis}
\end{figure}

\begin{figure}
    \centering
    \includegraphics[width=0.4\textwidth]{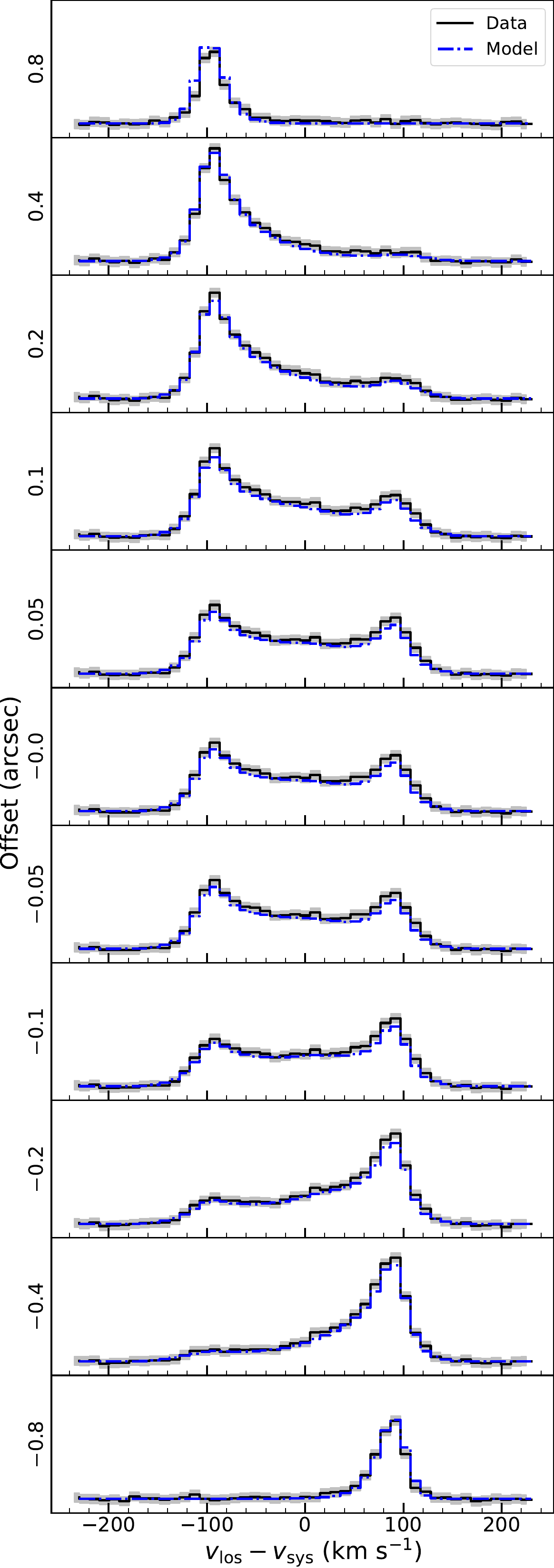}
    \caption{Observed and modeled \cotwo\ line profiles along the kinematic major axis of the NGC~3862 CND. In each panel, the observed spectrum is overlaid with the best-fit \kinms\ model. Position offsets from the kinematic center are indicated as the ordinates of each panel. The shaded gray region marks the $1\sigma$ RMS noise. All models reproduce the observed profiles across the CND's full radial extent.}
    \label{fig:specmap} 
\end{figure}

\begin{figure*}[!th]
    \centering
    \includegraphics[width=0.9\textwidth]{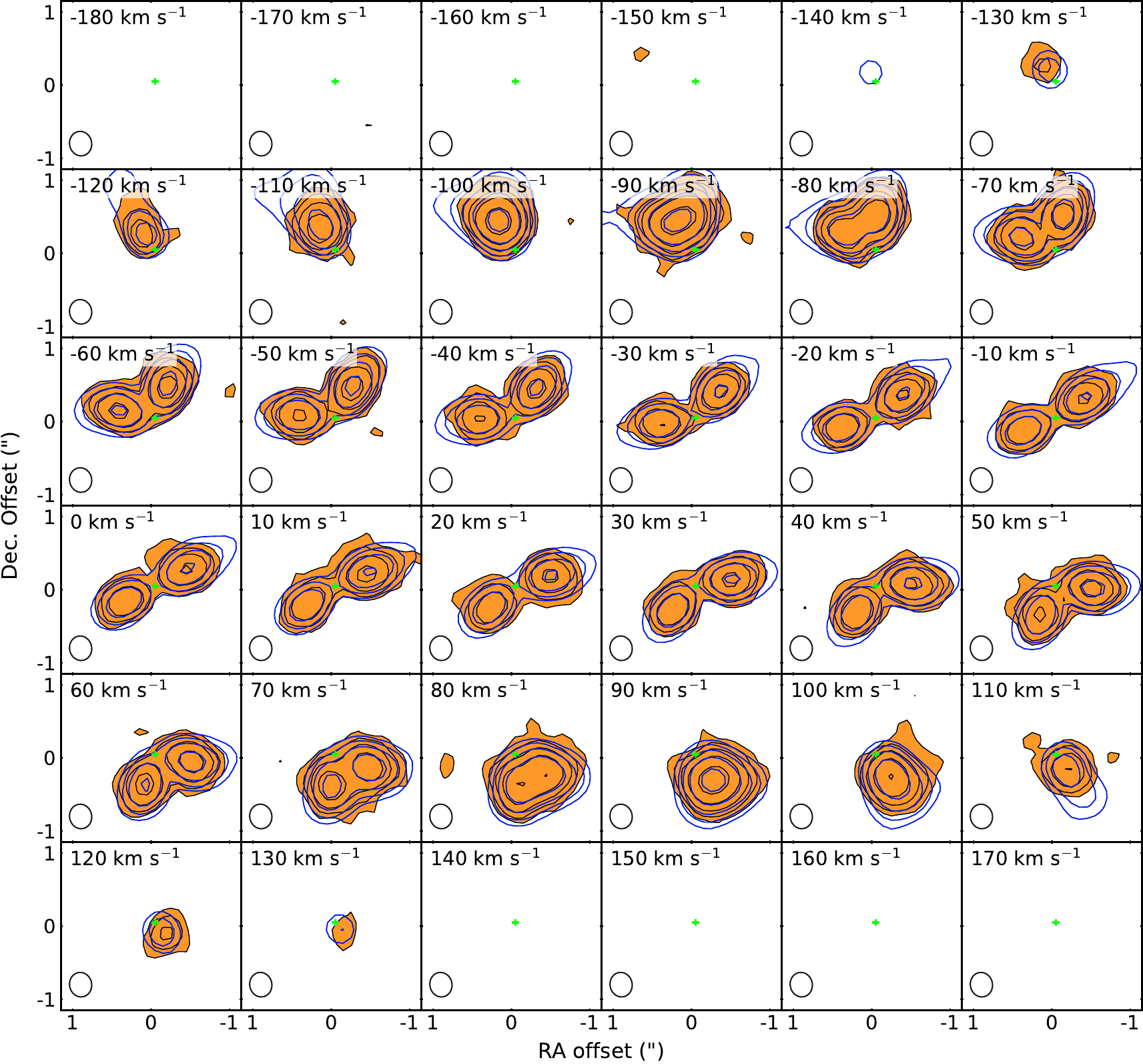}
    \caption{Channel maps of the \cotwo\ emission in NGC~3862. In each panel, the observed flux (orange contours) is overlaid with the best-fit \kinms\ model using the unmasked stellar-mass model (blue contours). The LOS velocity relative to $v_{\rm sys}$ is indicated in the upper-left corner of each panel. Both data and model are shown on the same logarithmic intensity scale. The synthesized beam is shown as a black ellipse in the lower-left corner. The model reproduces the spatial distribution and morphology of the observed emission across all velocity channels.}
    \label{fig:channels_unmasked} 
\end{figure*}

\begin{figure*}[!th]
\centering
    \includegraphics[width=0.98\textwidth]{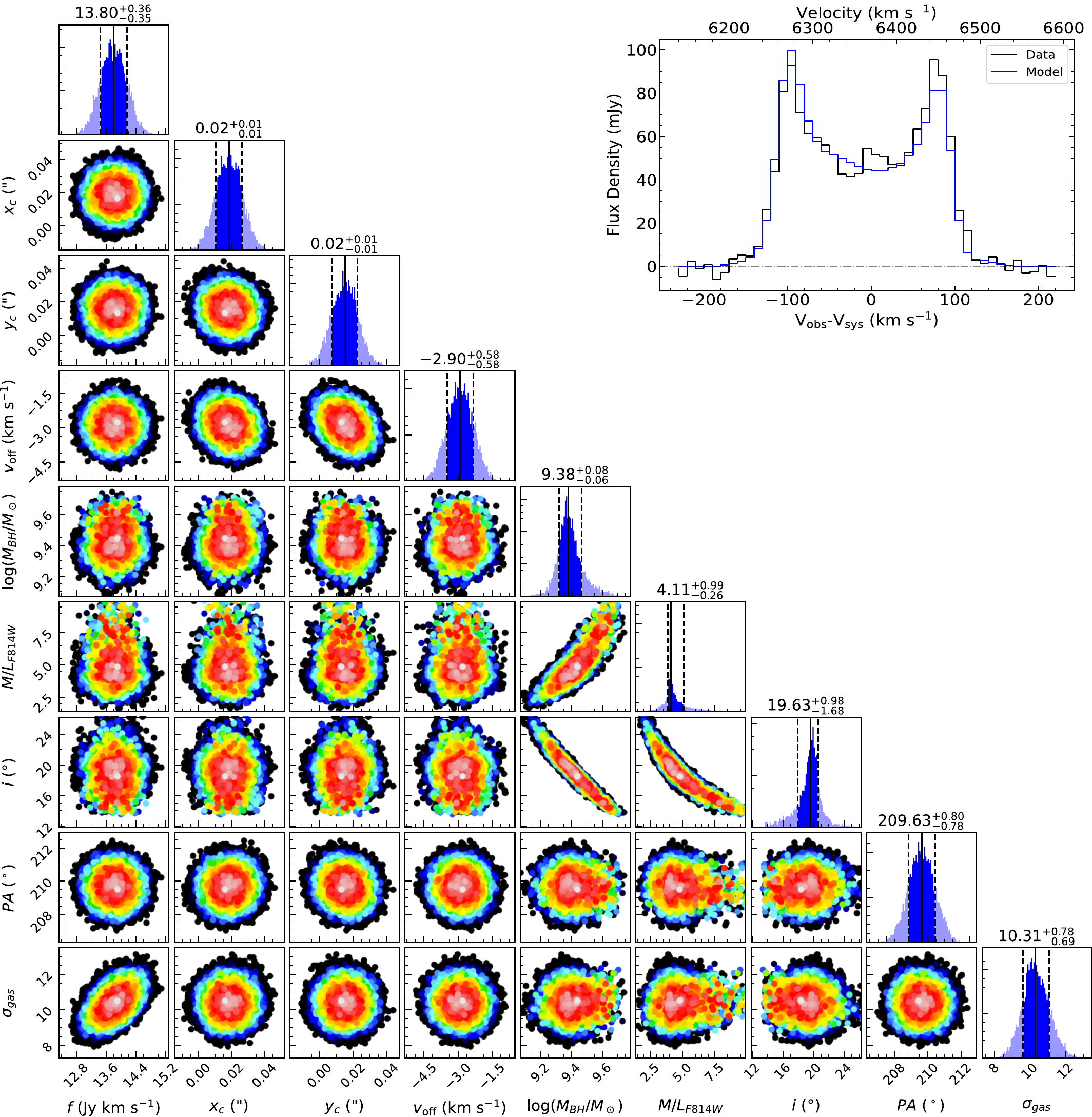}
    \caption{Corner plot of the \kinms\ MCMC posterior distributions for the best-fitting \kinms\ model. The diagonal panels show the 1D marginalized posterior distribution for each free parameter; vertical solid and dashed lines indicate the best-fit value and $1\sigma$ CL, respectively. The off-diagonal panels show the 2D marginalized distributions for each parameter pair, with contours enclosing the $1\sigma$ (white), $2\sigma$ (red), $3\sigma$ (green), and $>$$3\sigma$ (black) CLs. The best-fit parameter values and uncertainties are summarized in Table~\ref{tab:mcmc-results}.    {\it Insert panel:} Comparison of the observed \cotwo\ integrated spectrum (black; Panel~F of Figure~\ref{fig:moment-maps}) with the corresponding spectrum extracted from the best-fit \kinms\ model (blue), demonstrating good agreement between the data and model across the full velocity extent of the CND.} 
    \label{fig:MCMC_unmasked}
\end{figure*}

\subsection{Bayesian Inference and Priors} \label{sec:bayesian-infer}

We constrained the best-fit \kinms\ parameters and their uncertainties with the adaptive Metropolis Markov chain Monte Carlo (MCMC) algorithm \citep{Haario01}, implemented in the \textsc{AdaMet}\footnote{v2.0.9: \url{https://pypi.org/project/adamet/}} package \citep{Cappellari13a}. Each chain was run for $10^5$ iterations, discarding the first 20\% as burn-in. From the remaining samples we built the marginalized posteriors for each free parameter, took the best-fit values as the maximum-likelihood estimates, and report uncertainties at the $1\sigma$ (16th--84th percentile) and $3\sigma$ (0.13th--99.87th percentile) confidence levels (CL). \Mbh\ was sampled logarithmically and all other parameters linearly. We verified convergence by visual inspection of the chains. The prior ranges and initial guesses are listed in Table~\ref{tab:mcmc-results}.

The likelihood was evaluated over a $30 \times 30$ pixel region ($3\arcsec \times 3\arcsec$) and a velocity range of 460~\kms\ across 46 channels, for a total of $30 \times 30 \times 46 = 41{,}400$ data elements. Of these, $N = 4{,}348$ pixels contain significant \cotwo\ emission and constitute the constraints on the model.

In interferometric cubes such as ALMA's, the synthesized beam correlates adjacent pixels---the noise covariance \citep{Nguyen_2026_4061, Nguyen_2026_315}---so the formal per-pixel uncertainties underestimate the true statistical errors. We therefore rescaled $\sigma_{\rm RMS}$ by $(2N)^{0.25} \approx 10$, a correction for correlated noise \citep{vandenBosch09, Mitzkus17} that has been used in WISDOM \citep[e.g.,][]{North_2019, Smith19} and MBHBM$_\star$ \citep[e.g.,][]{Nguyen_2020, Nguyen_2022} studies and in VLT/MUSE analyses \citep{Thater_2022, Thater_2023}.

The log-likelihood is $\ln \mathcal{L} \equiv \ln p({\rm data} \mid {\rm model}) = -\tfrac{1}{2}\chi^2 + {\rm constant}$, with
\begin{equation}
\chi^2 \equiv \sum_j^N \frac{({\rm data}_j - {\rm model}_j)^2}{\sigma_j^2}
= \frac{1}{\sigma_{\rm RMS}^2} \sum_j^N ({\rm data}_j - {\rm model}_j)^2,
\label{eq:chi2}
\end{equation}
where the sum runs over all pixels with significant \cotwo\ emission in the fitting region, and $\sigma_{\rm RMS}$ is the per-channel noise measured from line-free channels (Section~\ref{sec:momentmaps}), taken to be spatially uniform.

\begin{table*}[!th]
\caption{Best-fitting \kinms\ parameters and their uncertainties}
\centering
\begin{tabular}{lcccccr}
\hline \hline
Model parameters&Search range& Initial guesses&Median&1$\sigma$ (16--84\%)&3$\sigma$ (0.14--99.86\%)\\
(1)   & (2)  &   (3)   &   (4)    &    (5)   & (6)\\
\hline
\underline{Mass model:}        & ~ & ~ & ~ & ~ & ~ \\
$\log(M_{\rm BH}$/\Msun)       &    6 $\rightarrow$ 11 & 9.38  &  9.38   & +0.08,$-$0.06 & +0.31,$-$0.24 \\
$M/L_{\rm F814W}$ (\Msun/\Lsun)&    0 $\rightarrow$ 10 & 4.05  & 4.11    & +0.99,$-$0.26 & +5.80,$-$2.10 \\ [1mm]
\underline{\cotwo\ CND:}       & ~ & ~ & ~ & ~ & ~ \\
$f$ (Jy \kms)                  & 1 $\rightarrow$ 50    & 13.8  & 13.80   & +0.36,$-$0.35 & +1.00,$-$0.98 \\
$i$ (\degr)                    &    0 $\rightarrow$ 90 & 20    & 19.63   & +0.98,$-$1.68 & +5.90,$-$5.90 \\ 
PA (\degr)                     & 0$\rightarrow$ 359    & 209   & 209.63  & +0.80,$-$0.78 & +2.30,$-$2.20 \\
\siggas (\kms)                 &0  $\rightarrow$ 20    & 10.50 &  10.31  & +0.78,$-$0.69 & +2.20,$-$2.00 \\ [1mm]
\underline{Nuisance:}          & ~ & ~ & ~ & ~ & ~ \\
$x_c$ (arcsec)                 &$-$0.5$\rightarrow$+0.5&   0   & 0.02 & +0.01,$-$0.01 & +0.02,$-$0.02 \\
$y_c$ (arcsec)                 &$-$0.5$\rightarrow$+0.5&   0   & 0.02 & +0.01,$-$0.01 & +0.02,$-$0.02 \\
$v_{\rm off}$ (\kms)           &$-$10$\rightarrow$+10  &$-$2.85&$-$2.90& +0.58,$-$0.58 & +1.60,$-$1.70 \\
\hline
\end{tabular}
\label{tab:mcmc-results}
\end{table*}

\subsection{Results}\label{sec:results}

Our fiducial \kinms\ model achieves a reduced chi-squared of $\chi^2_{\rm red} = 1.27$ and yields $M_{\rm BH} = 2.40^{+0.52}_{-0.32} \times 10^{9}$~\Msun\ and \ml$_{\rm F814W} = 4.11^{+0.99}_{-0.27}$~\Msun/\Lsun\ at the $1\sigma$ CL. This \ml$_{\rm F814W}$ agrees with stellar population synthesis (SPS) predictions of $M/L_I \sim 2$--5~\Msun/\Lsun\ for old ($\gtrsim$10~Gyr), metal-rich populations \citep[e.g.,][]{Maraston05, Bruzual03}, and with a Salpeter initial mass function (IMF), as found in massive ETGs of comparable luminosity and mass \citep[e.g.,][]{Cappellari2013a, Cappellari16}. This gives us confidence that the stellar mass model is physically reasonable and that \Mbh\ is not biased by an unrealistic \ml. The remaining best-fit parameters are listed in Table~\ref{tab:mcmc-results}.

Figure~\ref{fig:compare-pvd} compares the observed and modeled major-axis PVDs; the best-fitting \kinms\ model is shown in the middle panel. Two other models bracket this solution. The first assumes no SMBH ($M_{\rm BH} = 0$~\Msun) with \ml$_{\rm F814W} = 5.03$~\Msun/\Lsun: it reproduces the extended CND kinematics at $r > 0\farcs5$ but misses the central velocity rise of $\approx$40~\kms\ (left panel). The second fixes $M_{\rm BH} = 3.61 \times 10^9$~\Msun\ with \ml$_{\rm F814W} = 3.41$~\Msun/\Lsun: it matches the outer kinematics but overpredicts the central velocity at small radii (right panel). In both cases only \ml$_{\rm F814W}$ was free, with \Mbh\ fixed at the stated value and all other parameters held at their Table~\ref{tab:mcmc-results} best-fit values.

Figure~\ref{fig:momentmaps_compare} compares the observed (top row) and best-fit \kinms\ model (middle row) moment maps of the \cotwo\ CND. The zeroth-moment maps show that the model reproduces the ring-like intensity distribution, including the central depression and the nearly circular outer morphology. The first-moment maps show that it captures the regular, symmetric velocity field, with approaching (blue) and receding (red) emission reaching $\approx\pm130$~\kms\ relative to $v_{\rm sys}$. The second-moment maps agree well in the low-dispersion outer disk ($\sigma_{\rm LOS} \lesssim 20$~\kms) and reproduce the central enhancement elongated along the kinematic major axis, as expected from beam smearing.

The bottom-row panels show the normalized residuals, $({\tt Data} - {\tt Model})/\sigma$, for all three moments. The zeroth-moment residuals are mostly within $\pm5\sigma$ across the outer disk, but show a systematic pattern in the center: a positive ring ($\sim$$+2\sigma$) at $\approx$0\farcs2--0\farcs3 separating a negative core ($\sim$$-5\sigma$) from a negative annulus ($-0.5$ to $-2\sigma$) at $\approx$0\farcs5--0\farcs8. This pattern shows that the model captures the bulk emission but not the detailed radial profile of the CND. The central negative residual likely reflects beam smearing together with the sharp central depression, both hard to match with a spatially uniform gas model; the outer negative ring suggests the model slightly overpredicts the emission at intermediate radii, where the assumed azimuthal symmetry misses the asymmetric outer contours. These residuals are confined to the integrated intensity and do not affect \Mbh, which is constrained by the velocity field rather than the intensity distribution. The first- and second-moment residuals are mostly within $1.5\sigma$ with no large-scale structure, consistent with low-level non-circular motions or mild disk asymmetries at the noise level.

Specifically, the first-moment residual shows a mild kinematic twist on the eastern side of the disk, possibly a slight warp in the outer CND. The second-moment residual shows a ring of positive values near the nucleus, which may be a marginal underprediction of the central dispersion or residual beam smearing unresolved at our resolution. In both cases the residuals are small relative to the kinematic uncertainties and do not affect \Mbh.

We also checked for non-circular motions and kinematic warps (PA variations that would twist the isovelocity contours along the CND minor axis), which could bias the modeling and \Mbh. As shown in Figure~\ref{fig:PVD_minoraxis}, the minor-axis PVD of NGC~3862, extracted along PA~$= 209\degr + 90\degr$, is symmetric across all four ``forbidden quadrants'' and is well reproduced by the best-fit \kinms\ model, confirming the regular rotation of the \cotwo\ CND.

At the level of the data used in the likelihood, Figure~\ref{fig:specmap} shows the observed and modeled \cotwo\ line profiles at representative positions along the CND major axis, from the disk edge to the nucleus. The model reproduces the profiles across the full radial extent, including the innermost positions where the SMBH's influence is strongest. Figure~\ref{fig:channels_unmasked} shows the fit channel by channel, with close agreement between observed and modeled emission across the full velocity range of the CND.

Figure~\ref{fig:MCMC_unmasked} shows the marginalized 1D and 2D posteriors for all parameter pairs. In the off-diagonal panels, color indicates the likelihood, from white (maximum likelihood, $1\sigma$) to blue ($3\sigma$); the diagonal panels show the 1D posteriors, with solid lines marking the best-fit values and dashed lines the $1\sigma$ CL. Most posteriors are near-Gaussian; the exceptions are \Mbh, \ml$_{\rm F814W}$, and $i$, whose posteriors are asymmetric and mutually correlated through the \Mbh--$i$--\ml\ degeneracy of a nearly face-on disk.

The inset panel compares the observed and best-fit integrated \cotwo\ spectra. The model matches the double-horned profile and the overall flux distribution across the full velocity range of the CND.

Several parameter covariances appear in the posterior. The dominant one involves the inclination: because both \Mbh\ and the stellar mass scale as $1/\sin^2 i$, each is anti-correlated with $i$, so lowering $i$ raises both together. Through this shared dependence, \Mbh\ and \ml$_{\rm F814W}$ become positively correlated: the opposite of the usual anti-correlation seen in resolved data \citep[e.g.,][]{Davis_2013, Boizelle_2021}, where the two trade off against each other in the central potential. In a nearly face-on disk like NGC~3862 the $i$ degeneracy dominates and sets the sign of the \Mbh--\ml\ covariance. Two weaker covariances are also present: $y_{\rm cen}$ with $v_{\rm off}$, from the finite beam size when the kinematic center is tied to the continuum peak (Section~\ref{sec:cont})---higher angular resolution would localize the disk center and remove it---and $f$ with $\sigma_{\rm gas}$, since $f$ normalizes the adopted gas surface brightness \citep[e.g.,][]{Smith19}.

\begin{figure}
    \centering\includegraphics[width=0.9\columnwidth]{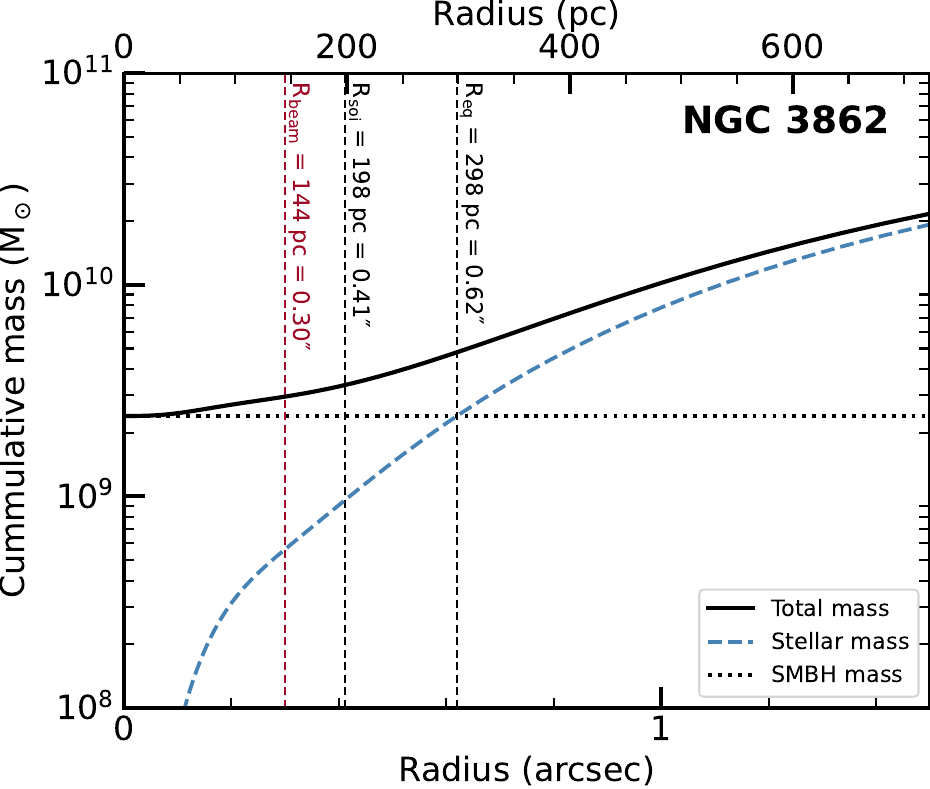}
    \caption{Enclosed mass of NGC~3862 as a function of radius: total (black solid), stellar (dashed), and SMBH (dotted). Vertical lines mark the synthesized beam ($R_{\rm beam} \approx 0\farcs30$), the classical SOI radius ($R_{\rm SOI} = 0\farcs41$), and the mass-based influence radius ($R_{\rm eq} = 0\farcs62$).}
    \label{fig:cumulative_mass}
\end{figure}

\subsection{Uncertainties} \label{sec:uncertainties}

Several systematic uncertainties affect our \Mbh\ measurement: (i) the adopted distance, (ii) the geometrically thin-disk assumption, (iii) the spatially uniform turbulent velocity dispersion, (iv) the AGN-contaminated component in the F814W MGE model, and (v) the dust-extinction correction of the F814W image.  Adding (ii), (iii), and (v) in quadrature gives a total systematic uncertainty of $\approx$38\% ($\pm0.91 \times 10^9$~\Msun) on \Mbh; the AGN test (iv) is illustrative only, and the distance uncertainty (i) is quoted separately with $\approx$+44\%/-35\%. We discuss each below.

\subsubsection{Distances} \label{sec:distances}
 
Since the inferred \Mbh\ scales linearly with the assumed angular-size distance ($M_{\rm BH} \propto D_A$), distance uncertainty constitutes a significant systematic in our \Mbh\ measurement. Published distance estimates for NGC~3862 span 65--143~Mpc (NED\footnote{\url{https://ned.ipac.caltech.edu}}), reflecting the absence of a primary distance indicator for this galaxy. We adopt the distance of $D = 99.4$~Mpc from \citet{Ma_2014}, derived from the galaxy redshift assuming our adopted $\Lambda$CDM cosmology (Section~\ref{sec:intro}). The full range of published distances introduces a systematic uncertainty of $\approx$+44\%/-35\% on \Mbh.

\subsubsection{Thick Disk Assumption} \label{sec:thickdisk}

Our fiducial \kinms\ model assumes a geometrically thin \cotwo\ CND, with the vertical scale height fixed to zero. To test this, we refit with the scale height $d$ free, taking the gas density uniform in the vertical direction within the disc. The best-fit parameters change by less than 5\% from the fiducial values (Table~\ref{tab:mcmc-results}), and the inferred $d = 0\farcs06 \pm 0\farcs04$ is well below the beam size ($0\farcs3$), so the thin-disk assumption has a negligible effect on \Mbh.

\begin{figure}[!th]
    \centering
    \includegraphics[width=0.9\columnwidth]{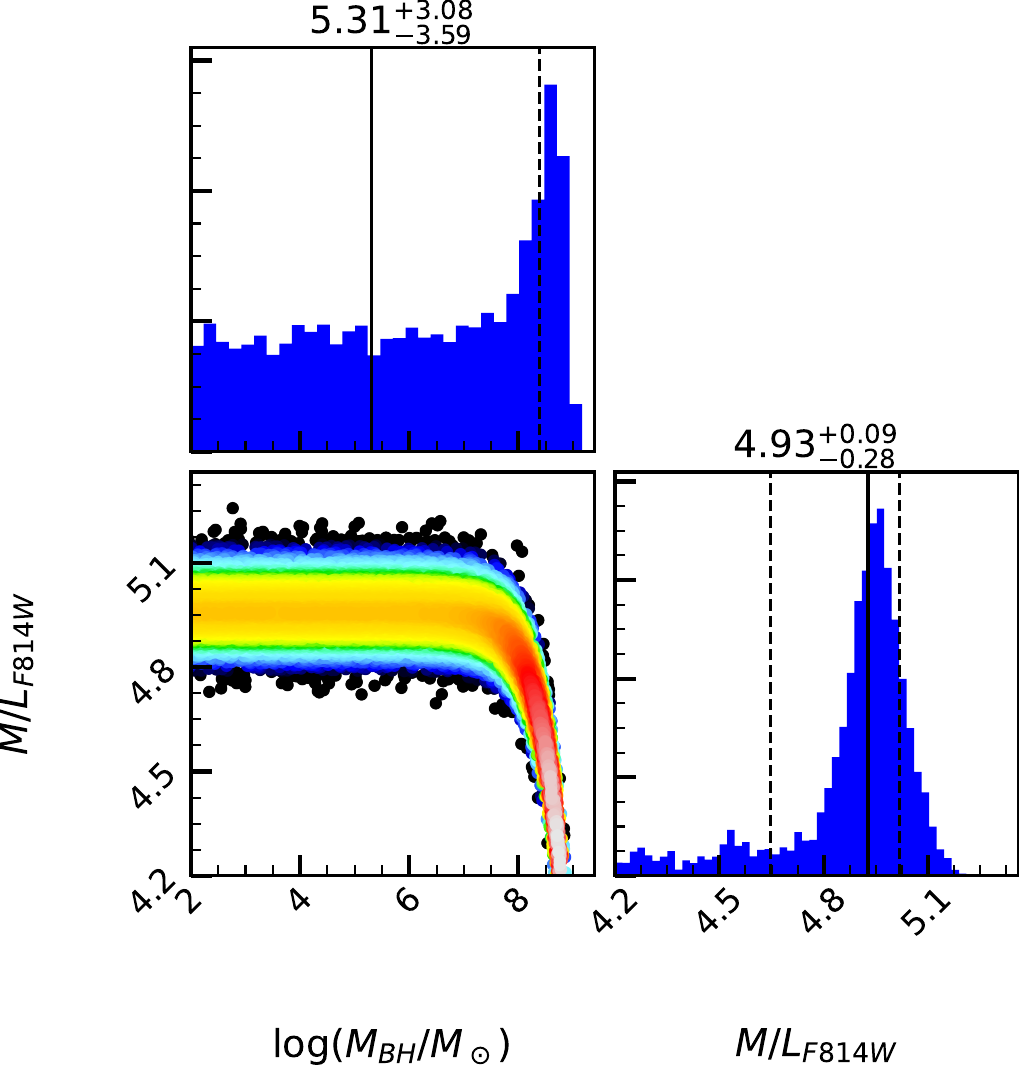} 
    \caption{Posterior of $\log(M_{\rm BH}/M_\odot)$ and $M/L_{\rm F814W}$ when the innermost Gaussian---treated as the AGN component in the fiducial model---is kept in the stellar MGE, with all other parameters fixed at their best-fit values (Table~\ref{tab:mcmc-results}). The \Mbh\ posterior is flat toward low masses and consistent with zero, while $M/L_{\rm F814W}$ remains well constrained near its fiducial value (Section~\ref{sec:agn}): mis-attributing the AGN point source to starlight adds a central point mass that substitutes for the black hole and washes out the \Mbh\ measurement.}
    \label{fig:bhmass-ml-AGN}
\end{figure}

\subsubsection{Turbulent Velocity Dispersion of the Gas} \label{sec:sigma}

Our fiducial \kinms\ model assumes a spatially uniform $\sigma_{\rm gas}$, but in practice it may vary with radius and azimuth, and beam smearing can inflate the observed central dispersion and bias \Mbh. To test this, we refit with $\sigma_{\rm gas}$ parameterized in three ways, each with a floor $\sigma_{\rm gas,min} = 1$~\kms\ to avoid unrealistically narrow line profiles \citep{Barth16b}.

\textit{Linear:} $\sigma_{\rm gas}(r) = \alpha r + \beta$. The fit gives $\alpha \approx 0$ and $\beta = 10.71$~\kms, i.e.\ no significant radial gradient, with all other parameters consistent with the fiducial model (Table~\ref{tab:mcmc-results}).

\textit{Exponential:} $\sigma_{\rm gas}(r) = \sigma_0 \exp(-r/r_0) + \sigma_1$, with $\sigma_0$, $\sigma_1$, $r_0$ free. This gives $M_{\rm BH} = 2.31^{+0.09}_{-0.16} \times 10^9$~\Msun, \ml$_{\rm F814W} = 4.12 \pm 0.02$~\Msun/\Lsun, $\sigma_0 = 46.1^{+5.9}_{-3.6}$~\kms, $\sigma_1 = 10.6^{+0.6}_{-0.4}$~\kms, and $r_0 = 0\farcs03 \pm 0\farcs02$.

\textit{Gaussian:} $\sigma_{\rm gas}(r) = \sigma_0 \exp[-(r - r_0)^2/2\mu^2] + \sigma_1$, with $\sigma_0$, $\sigma_1$, $\mu$, $r_0$ free ($r_0$ allowed either sign, to allow the peak dispersion to be offset from the center). This gives $M_{\rm BH} = 2.42^{+0.11}_{-0.09} \times 10^9$~\Msun, \ml$_{\rm F814W} = 4.09^{+0.02}_{-0.04}$~\Msun/\Lsun, $\sigma_0 = 36.7^{+3.7}_{-5.7}$~\kms, $\sigma_1 = 10.2^{+1.1}_{-0.7}$~\kms, $r_0 = 0\farcs02^{+0\farcs04}_{-0\farcs05}$, and $\mu = 0\farcs07 \pm 0\farcs03$.

All three \Mbh\ values agree with the fiducial constant-$\sigma_{\rm gas}$ result (Section~\ref{sec:results}): the linear gradient leaves \Mbh\ essentially unchanged, and the exponential and Gaussian profiles shift it by $\lesssim$4\%. \Mbh\ is therefore insensitive to the assumed $\sigma_{\rm gas}(r)$.

\begin{figure*}[!th]
    \centering
    \includegraphics[width=0.98\textwidth]{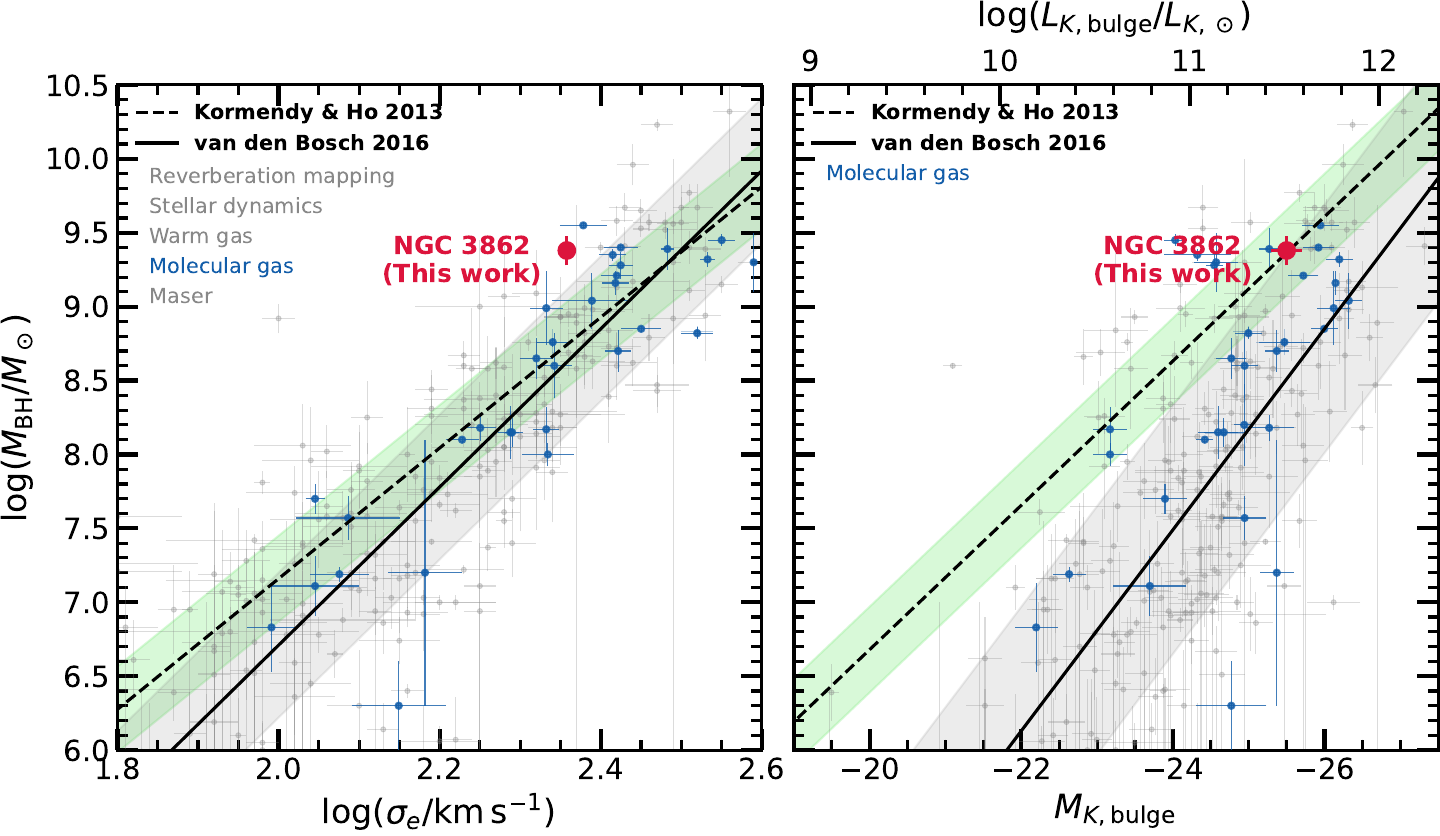}
    \caption{Molecular-gas \Mbh\ for NGC~3862 (red) compared with the \Mbh--$\sigma_e$ (left) and \Mbh--$L_{K,\rm bulge}$ (right) relations of \citet{Kormendy13} (dashed) and \citet{Bosch_2016} (solid); shaded bands show the $1\sigma$ intrinsic scatter of each relation. Literature measurements are shown in gray by method and in blue for molecular-gas determinations, as labeled.}
    \label{fig:bhmass-sigma}
\end{figure*}

\subsubsection{HST/F814W MGE Model with AGN Contamination}\label{sec:agn}

To gauge how AGN contamination of the stellar-mass model biases \Mbh, we refit with the innermost MGE Gaussian---the one flagged as the AGN component (Table~\ref{table:mge-table})---restored to the stellar model, letting only \Mbh\ and \ml$_{\rm F814W}$ vary and holding all else at the fiducial values (Table~\ref{tab:mcmc-results}). This gives $\log(M_{\rm BH}/M_\odot) = 5.31^{+3.08}_{-3.59}$ and \ml$_{\rm F814W} = 4.93^{+0.09}_{-0.28}$~\Msun/\Lsun. The posterior (Figure~\ref{fig:bhmass-ml-AGN}) is flat toward low masses and consistent with $M_{\rm BH} = 0$: attributing the AGN emission to stellar light inflates the central stellar mass enough to reproduce the \cotwo\ kinematics on its own, leaving \Mbh\ unconstrained. The best-fit \ml$_{\rm F814W}$ stays within the plausible range for an old, massive ETG, so the degeneracy is absorbed almost entirely by \Mbh\ rather than \ml, the spurious AGN mass substitutes directly for the black hole. We therefore use this fit only to illustrate how AGN contamination can turn \Mbh\ unmeasurable, not as a reliable measurement.

\subsubsection{Dust Extinction Correction of the F814W Image}\label{sec:dust-sys}

To assess the systematic from the dust-extinction correction (Section~\ref{sec:dust}), we refit using an MGE built from the uncorrected (dust-affected) image, again letting only \Mbh\ and \ml$_{\rm F814W}$ vary and holding all else at the fiducial values (Table~\ref{tab:mcmc-results}). This gives $\log(M_{\rm BH}/M_\odot) = 9.17$ and \ml$_{\rm F814W} = 3.99$~\Msun/\Lsun, lower than the fiducial $\log(M_{\rm BH}/M_\odot) = 9.38$ and \ml$_{\rm F814W} = 4.11$~\Msun/\Lsun\ by 38\% and 3\%, respectively. The dust correction is thus the dominant systematic in our \Mbh\ budget.

\subsection{The Reliability of Our Measurements}\label{sec:reliability}

At the adopted distance of $D=99.4$~Mpc, the classical SOI radius $R_{\rm SOI} = GM_{\rm BH}/\langle\sigma_\star\rangle_e^2 \approx 0\farcs41$ (198~pc; $\langle\sigma_\star\rangle_e = 228$~\kms, \citealt{Veale_2017}) exceeds the synthesized beam ($0\farcs3$) by a factor of $\approx$1.4. The mass-based influence radius of \citet{Merritt2004}, enclosing a stellar mass equal to \Mbh, is $R_{\rm eq} \approx 0\farcs62$ (299~pc), a factor of $\approx$2.1 larger. Following \citet{Boizelle_2019, Boizelle_2021}, who find $R_{\rm eq}$ a more appropriate measure of the SMBH gravitational domain in massive ETGs, we adopt $R_{\rm SOI} = R_{\rm eq} = 0\farcs62$ for the resolving-power assessment. The criterion $\zeta = 2R_{\rm SOI}/\theta_{\rm beam} \approx 4.1$ and its inclination-corrected value $\zeta\cos i \approx 3.9$ both exceed the threshold $\zeta = 2$ for a reliable dynamical \Mbh\ measurement \citep{Boizelle_2019, Nguyen_2020}, so our ALMA observations spatially resolve the SMBH SOI in NGC~3862.

\subsection{\Mbh\ on the Scaling Relations}\label{sec:scaling_relations}

The only previous \Mbh\ estimate for NGC~3862 is an upper limit, \Mbh $\leq 6\times10^8$~\Msun\ \citep{Beifiori_2012}, from HST/STIS ionized-gas line widths, a factor of $\approx$4 below our dynamical value. This is not surprising: the \Mbh--$\sigma_e$ relation has an intrinsic scatter of $\sim$0.29~dex \citep{Kormendy13}, and massive ETGs at the high-mass end tend to host overmassive SMBHs relative to the mean relations \citep[e.g.,][]{McConnell2011, Thomas2016, Mehrgan2019, Nguyen_2026_highz, Cappellari2026_JWST2, Cappellari2026_JWST3}. In the \Mbh--$\sigma_e$ plane (Figure~\ref{fig:bhmass-sigma}, left), NGC~3862 sits $\approx$0.6-0.8~dex above the \citet{Kormendy13} and \citet{Bosch_2016} relations---about $1.5-2.2\sigma$ of the intrinsic scatter, at the upper edge of the scatter band. In the \Mbh--$L_{K,\rm bulge}$ plane (right), it lies on the \citet{Kormendy13} relation but $\approx$0.9~dex ($\approx$1.0$\sigma$) above that of \citet{Bosch_2016}, consistent with an overmassive SMBH for its bulge luminosity. NGC~3862 thus appears overmassive in both scaling relations, more so against $L_{K,\rm bulge}$ than against $\sigma_e$, as expected for a massive core elliptical at the high-mass end \citep[e.g.,][]{Nguyen_2023, de_Nicola_2025}.

\section{Conclusions}\label{sec:conclusion}

Our \kinms\ modeling of the ALMA \cotwo\ CND kinematics in NGC~3862, obtained at $\theta_{\rm beam} \approx 0\farcs3$ (144~pc) in Cycle~7, yields a SMBH mass $M_{\rm BH} = (2.40^{+0.52}_{-0.32}\,[\rm stat.] \pm 0.91\,[\rm syst.]) \times 10^9$~\Msun\ and a stellar mass-to-light ratio \ml$_{\rm F814W} = 4.11^{+0.99}_{-0.27}\,[\rm stat.] \pm 0.17\,[\rm syst.]$~\Msun/\Lsun\ (Section~\ref{sec:uncertainties}). This \ml$_{\rm F814W}$ is consistent with a single, old ($\gtrsim$10~Gyr) stellar population, in line with the uniform stellar F475W$-$F814W $\approx 2$~mag color away from the jet and dust lane.

\section*{Acknowledgements} 
This paper makes use of the following ALMA data: ADS/JAO.ALMA\#2019.1.00036.S. ALMA is a partnership of ESO (representing its member states), NSF (USA), and NINS (Japan), together with NRC (Canada) and NSC and ASIAA (Taiwan), and KASI (Republic of Korea), in cooperation with the Republic of Chile. The Joint ALMA Observatory is operated by ESO, AUI/NRAO, and NAOJ. The National Radio Astronomy Observatory is a facility of the National Science Foundation operated under cooperative agreement by Associated Universities, Inc.

\facility{ALMA and \hst/WFC3}

\software{{\tt Python~v3.12} \citep{VanRossum2009}, 
{\tt Matplotlib~v3.6} \citep{Hunter2007}, 
{\tt NumPy~v1.22} \citep{Harris2020}, 
{\tt SciPy~v1.3} \citep{Virtanen2020},  
{\tt photutils~v0.7} \citep{bradley2024}, 
{\tt AstroPy~v5.1} \citep{AstropyCollaboration2022}, 
{\tt AdaMet~v2.0} \citep{Cappellari2013a}, 
{\tt JamPy~v7.2} \citep{Cappellari2020}, and
{\tt MgeFit~v5.0} \citep{Cappellari2002}.
}

\bibliographystyle{aasjournalv7}
\bibliography{references} 


\label{lastpage}
\end{document}